\RequirePackage{fix-cm}
\documentclass[natbib,smallextended]{svjour3}       
\smartqed  
\usepackage{graphicx}
\usepackage{amssymb}
\usepackage{aas_macros}
\usepackage[colorlinks=true,urlcolor=blue,citecolor=blue,linkcolor=blue,breaklinks]{hyperref}
\def\vkm{km s$^{-1}$}
\def\vkme{\textrm{km s}^{-1}}
\def\eqt#1#2{#1_\textrm{\scriptsize #2}}                                                                                
\def\degree{$^\circ$}

\def\arcsa#1#2{$#1^{\prime\prime}_{^\textrm{.}}#2$}

\newcommand{\solarmass}{\ensuremath{\, M_\odot}}
\newcommand{\solarlum}{\ensuremath{\, L_\odot}}

\def\cmc{cm$^{-3}$}

\def\micron{$\mu$m}

\def\mH2{m_{\textrm{\scriptsize H}_2}}

\def\H2{H$_2$}
\def\N2HP{N$_2$H$^+$}

\def\NH3{NH$_3$}

\journalname{The Astronomy and Astrophysics Review}
\begin{document}

\title{Molecular jets from low-mass young protostellar objects}

\author{Chin-Fei Lee}

\institute{Academia Sinica Institute of Astronomy and Astrophysics,
P.O. Box 23-141, Taipei 106, Taiwan \email{cflee@asiaa.sinica.edu.tw}
\and
Graduate Institute of Astronomy and Astrophysics, National Taiwan
   University, No.\ 1, Sec.\ 4, Roosevelt Road, Taipei 10617, Taiwan}

\date{Received: date / Accepted: date}

\maketitle

\begin{abstract}
Molecular jets are seen coming from the youngest protostars in the early
phase of low-mass star formation.  They are detected in CO, SiO, and SO at
(sub)millimeter wavelengths down to the innermost regions, where their
associated protostars and accretion disks are deeply embedded and where they
are launched and collimated.  They are not only the fossil records of
accretion history of the protostars but also are expected to play an
important role in facilitating the accretion process.  Studying their
physical properties (e.g., mass-loss rate, velocity, rotation, radius,
wiggle, molecular content, shock formation, periodical variation, magnetic
field, etc) allows us to probe not only the jet launching and collimation,
but also the disk accretion and evolution, and potentially binary formation
and planetary formation in the disks.  Here I review recent exciting results
obtained with high-spatial and high-velocity resolution observations of
molecular jets in comparison to those obtained in the optical jets in the
later phase of star formation.  Future observations of molecular jets with a
large sample at high spatial and velocity resolution with ALMA are
expected to lead to a breakthrough in our understanding of jets
from young stars.


\end{abstract}

\setcounter{tocdepth}{3}
\tableofcontents

\section{Introduction}

Protostellar jets are spectacular signposts of star formation.  They are
highly collimated structures consisting of a chain of knots and bow shocks,
emanating from young protostellar objects, propagating away at highly
supersonic speeds.  Since they are believed to be launched from accretion
disks around protostars, they are not only the fossil records of accretion
history of the protostars but also are expected to play an important role in
facilitating the accretion process.  Studying their dynamics and evolution
also allows us to probe disk evolution and potentially planetary formation
in the disks.

In low-mass star formation, the protostellar phase with active accretion can
be divided into the Class 0 and Class I phases
\citep{Mckee2007,Evans2009,Kennicutt2012}.  The Class 0 phase starts when a
protostar is first formed at the center with a mass of $\sim 10^{-3}
\solarmass$ and a radius of $\sim 2\, R_\odot$
\citep{Larson1969,Masunaga2000}.  In this phase, the protostar is deeply
embedded in a large cold dust envelope and actively accreting material from
it.  This phase will end in $\sim 10^5$ yrs when the cold envelope decreases
to $\sim$ 10\% of its original amount \citep{Evans2009}.  This phase is then
followed by the Class I phase, which will end in $\sim 5\times 10^5$ yrs
when most of the envelope material has been consumed and the accretion
almost comes to an end \citep{Evans2009}.  Then the protostar enters the
Class II phase or T-Tauri phase, and becomes a pre-main-sequence star
visible in the optical.  Note that some protostars could already become
pre-main-sequence stars earlier during the Class I phase.

\begin{figure}[htb]
\centering
\includegraphics[angle=0, width=\textwidth]
       {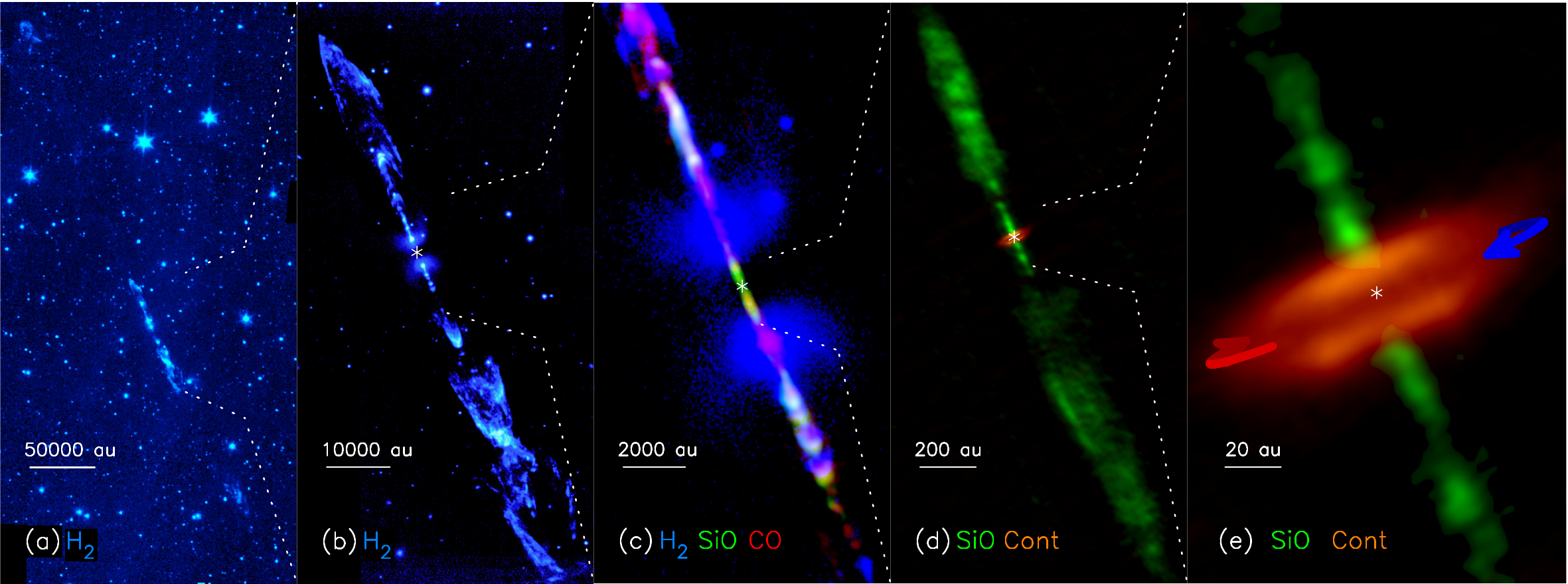}
\caption{HH 212: (a) H$_2$ map of the jet in parsec scale adopted from
\citet{Reipurth2019}.  (b) H$_2$ map of the inner jet adopted from
\citet{McCaughrean2002}.  (c) A composite image of the jet within $\sim$
10000 au of the central source, with H$_2$ map (blue) from
\citet{McCaughrean2002} and SiO (green) and CO (red) maps from
\citet{Lee2015}.  (d) The jet in SiO (green) within $\sim$ 1000 au of the central source
and the accretion disk in dust continuum at 850 \micron{}
(orange), adopted from \citep{Lee2017Jet}.  (e) shows the SiO jet (green)
within $\sim$ 100 au of the central source with the dusty accretion disk
(orange), adopted from \citep{Lee2017Jet}.  The blue and red arrows show the
disk rotation.
\label{fig:HH212_Jet}}
\end{figure}

Jets are commonly seen in the Class 0 to the early Class II phase when the
accretion process is active.  Their velocities scale with protostellar mass
and thus increase with the evolutionary phase, increasing from $\sim$ 100
\vkm{} in the early phase to a few 100 \vkm{} in the late phase
\citep{Hartigan2005,Anglada2007,Hartigan2011}.  Therefore, they can
propagate for a long distance into the ISM, producing spectacular
parsec-scale Herbig-Haro objects \citep{Reipurth2001} even in the Class 0
phase \cite[][see also Fig.~\ref{fig:HH212_Jet} for the Class 0 jet HH
212]{Reipurth2019}.  When these jets propagate into the ISM, they push and
sweep up the surrounding material, forming molecular outflows around the jet
axis \citep{Lee2000,Arce2007}, perturbing the ISM and thus potentially
reducing the star-formation efficiency \citep{Bally2016}.  Interestingly,
jets are also detected around brown dwarfs
\cite[e.g.,][]{Whelan2005,Riaz2017}, intermediate-mass protostars
\cite[e.g.,][]{Zapata2010,Reiter2017,Takahashi2019}, and high-mass
protostars
\citep{Reipurth2001,Zapata2006,Carrasco2010,Ellerbroek2013,Caratti2015},
suggesting that the low-mass star formation scenario may apply to these
objects.

Current observations also show an evolution in the gas content  of the jets. 
In the Class 0 phase, the jets are mainly detected in molecular gas, e.g.,
CO, SiO, and SO at (sub)millimeter wavelengths and H$_2$ in infrared
\citep{Arce2007,Frank2014}.  In the Class I and Class II phases, they are
mainly detected in atomic and ionized gas, e.g., O, H$\alpha$, and S II
\citep{Reipurth2001,Bally2016}, appearing as Herbig-Haro flows at optical
and infrared wavelengths.  At the base near the launching points, the jets
are ionized and thus radiate free-free emission at centimeter wavelengths
\citep{Anglada2018}.  Therefore, different telescopes are used to study
different regions of the jets in different evolutionary phases.


Recent comprehensive reviews of the jets have been presented by
\citet{Frank2014} and \citet{Bally2016}, and a further review on radio jets
by \citet{Anglada2018}.  Here I will focus on molecular jets seen in the
early phase of low-mass star formation, probing the initial phase of the jet
formation and accretion process in the first $10^{5}$ yrs in the Class 0
phase.  Without suffering from dust extinction effects, molecular lines at
(sub)millimeter wavelengths allow us to probe the jets close to the
innermost regions, where the sources are deeply embedded and where the jets
are launched and collimated (see Fig.~\ref{fig:HH212_Jet}).  The jets are
believed to be launched from accretion disks around protostars through
magneto-centrifugal forces \citep{Shu2000,Konigl2000}, and are thus expected
to be magnetized and rotating.  In this case, the jets can also solve the
angular momentum problem in the innermost edges of the disks by carrying away
angular momentum from there to allow disk material to fall onto the central
protostars.  I will review current exciting and revolutionary results on
molecular jets, especially those obtained with unprecedented angular
resolution, velocity resolution, and sensitivity using ALMA.  Together with
previous detailed studies of jets in their later phase, we can set strong
constraints on jet launching and collimation.  With a detailed study of jet
physical properties (e.g., mass-loss rate, velocity, rotation, radius,
wiggle, molecular content, shock formation, periodical variation, magnetic
field, etc), we can also probe accretion process, binary formation, disk
instability and evolution, and potentially planetary formation in the disks.

\section{Observed properties of molecular jets}



Protostellar jets in the Class 0 phase are mainly molecular.  They can be
traced by high-velocity CO emission
\citep{Gueth1999,Lee2007HH212,Santiago2009,Hirano2010,Plunkett2015}, which
allows us to derive the density and thus the mass-loss rate in the jets. 
Their knots and bow shocks can be traced by shock tracers, e.g., H$_2$
\citep{McCaughrean1994,Zinnecker1998}, SiO
\citep{Gueth1998,Hirano2006,Palau2006,Codella2007,Lee2007HH212,Codella2014,Podio2016,Bjerkeli2019},
and SO \citep{Lee2007HH212,Lee2010HH211,Codella2014}.  More importantly, SiO
is a dense shock tracer, tracing uniquely the jets within $\sim 10^4$ au
of the central sources down to the bases where the density is high, allowing
us to probe the jet launching and collimation regions.

\begin{table} 
\small 
\centering
\caption{Molecular jets from low-mass Class 0 protostars detected in both SiO and CO} 
\label{tab:jetsource} 
\begin{tabular}{llllllllll} 
\hline Source & $D$ & $L_{\mathrm{bol}}$ & $M_\ast$ & $r_d$ & $v_j$ & $\dot{M}_j$ & $L_j$ & $\dot{M}_{\mathrm{acc}}$ & Refs. \\
       & (pc) & ($L_\odot$) & ($M_\odot$)  & (au) & (km/s) & (\solarmass \ yr$^{-1}$)& $L_\odot$ & (\solarmass \ yr$^{-1}$) \\ \hline\hline
IRAS 04166+2706 & 140 & 0.4 & & ? & 61 & $0.7\times10^{-6}$ & 0.21 & & 1 \\ 
B335  & 100 & 0.7 & 0.05 & $<5$ &160 & $0.1\times10^{-6}$ & 0.21 & $1.2\times10^{-6}$ & 2\\ 
NGC1333 IRAS4A2 & 320 & 3.5 & 0.11 & $<65$ & 100 & & & & 3 \\ 
L1157  & 250 & 3.6 & 0.04 & $<50$ & 112 & $0.8\times10^{-6}$ & 0.82 & $7.1\times10^{-6}$ & 4\\ 
HH 211  & 320 & 4.7 & 0.08 & 16 & 100 & $1.1\times10^{-6}$ & 1.1 & $4.7\times10^{-6}$ & 5\\ 
L1448 C & 320 & 8.4 & 0.08 & $50?$ &160 & $2.4\times10^{-6}$ & 5.0 & $11\times10^{-6}$ & 6 \\ 
HH 212 & 400 & 9.0 & 0.25 & 44 &135 & $1.3\times10^{-6}$ & 1.9 & $2.8\times10^{-6}$  & 7\\ 
\hline
\multicolumn{10}{p{11cm}}{
Here $D$ is distance, $L_{\mathrm{bol}}$ is bolometric luminosity, $M_\ast$ is mass of central protostar,
$r_d$ is disk radius, $v_j$ is jet velocity, $\dot{M}_j$ is mass-loss rate in the jet, $L_j$ is mechanical
luminosity of the jet, and $\dot{M}_{\mathrm{acc}}$ is accretion rate.
References: (1) \citet{Santiago2009}, \citet{Tafalla2017} 
(2) \citet{Yen2010}, \citet{Green2013}, \citet{Bjerkeli2019}
(3) \citet{Choi2006,Choi2010,Choi2011} 
(4) \citet{Green2013}, \citet{Kwon2015}, \citet{Podio2016}, \citet{Maury2019} 
(5) \citet{Froebrich2005}, \citet{Lee2018HH211}, \citet{Jhan2016}
(6) \citet{Hirano2010}, \citet{Green2013}, \citet{Maury2019} 
(7) \citet{Zinnecker1998}, \citet{Lee2017Disk,Lee2017Jet} 
}   \\
\end{tabular}
\end{table}

Table~\ref{tab:jetsource} lists the properties of a few well-studied
molecular jets from low-mass Class 0 protostars detected in both SiO and CO at
(sub)millimeter wavelengths.  Note that although the jets listed here are
bipolar, there are also monopolar jets, e.g., in NGC1333-IRAS2A
\citep{Codella2014}.  This table also lists the mechanical luminosity of
the jets calculated from the following equation

\begin{equation} L_j =
\frac{1}{2}\dot{M}_j v_j^2 \approx 0.82 \ \frac{\dot{M}_j}{10^{-6}
\solarmass \ \textrm{yr}^{-1}} \left(\frac{v_j}{100 \; \vkme}\right)^2 L_\odot
\end{equation} 

\noindent and the accretion rate onto the protostars estimated from the
following equation

\begin{equation} 
\dot{M}_{\mathrm{acc}} \sim
\frac{(L_{\mathrm{bol}}+L_j)R_\ast}{G M_\ast} \approx 6.44 \times 10^{-8} \ 
\frac{L_{\mathrm{bol}}+L_j}{L_\odot} \frac{\solarmass}{M_\ast} \ \solarmass \  
\textrm{yr}^{-1} 
\end{equation} 

\noindent assuming that both the bolometric luminosity and the jet's mechanical
luminosity come from accretion.  Here we assume the radius of the central
protostars is $R_\ast \sim 2 \, R_\odot$ \citep{Stahler1988}.  Rotating disks have
been detected or suggested in most of these sources, as required in current
magneto-centrifugal models of jet launching \citep{Shu2000,Konigl2000}.

Although the sample is small and the measurements have significant
uncertainties, we still can see some trends in comparison to the jets in the
later phase of star formation.  The mass of the protostars is $0.05$--$0.25
\solarmass$, smaller than that in the Class I and II phases
\cite[see, e.g.,][]{Simon2000,Yen2017}.  Thus, the mass in most of the protostars
here might be too small to have deuterium burning \citep{Stahler1988}.  The
jet velocity is 61--160 \vkm{}, much smaller than that of the Class I and II
jets.  The mass-loss rate is (0.7-- 2.4)$\times10^{-6} \solarmass \mathrm{\
yr}^{-1}$ (except for B335 with a much smaller value), and is much larger
than those in the Class I and II phases.  The jets have a mechanical
luminosity about 20--50\% of the bolometric luminosity.  The accretion rate
is a few times the mass-loss rate.  The disks and candidate disks mostly
have a radius $r_d <$ 50 au, smaller than those in the Class I phase, as
discussed in \citet{Maury2019}.

\begin{figure}[htb]
\centering
\includegraphics[angle=270, width=\textwidth]
                {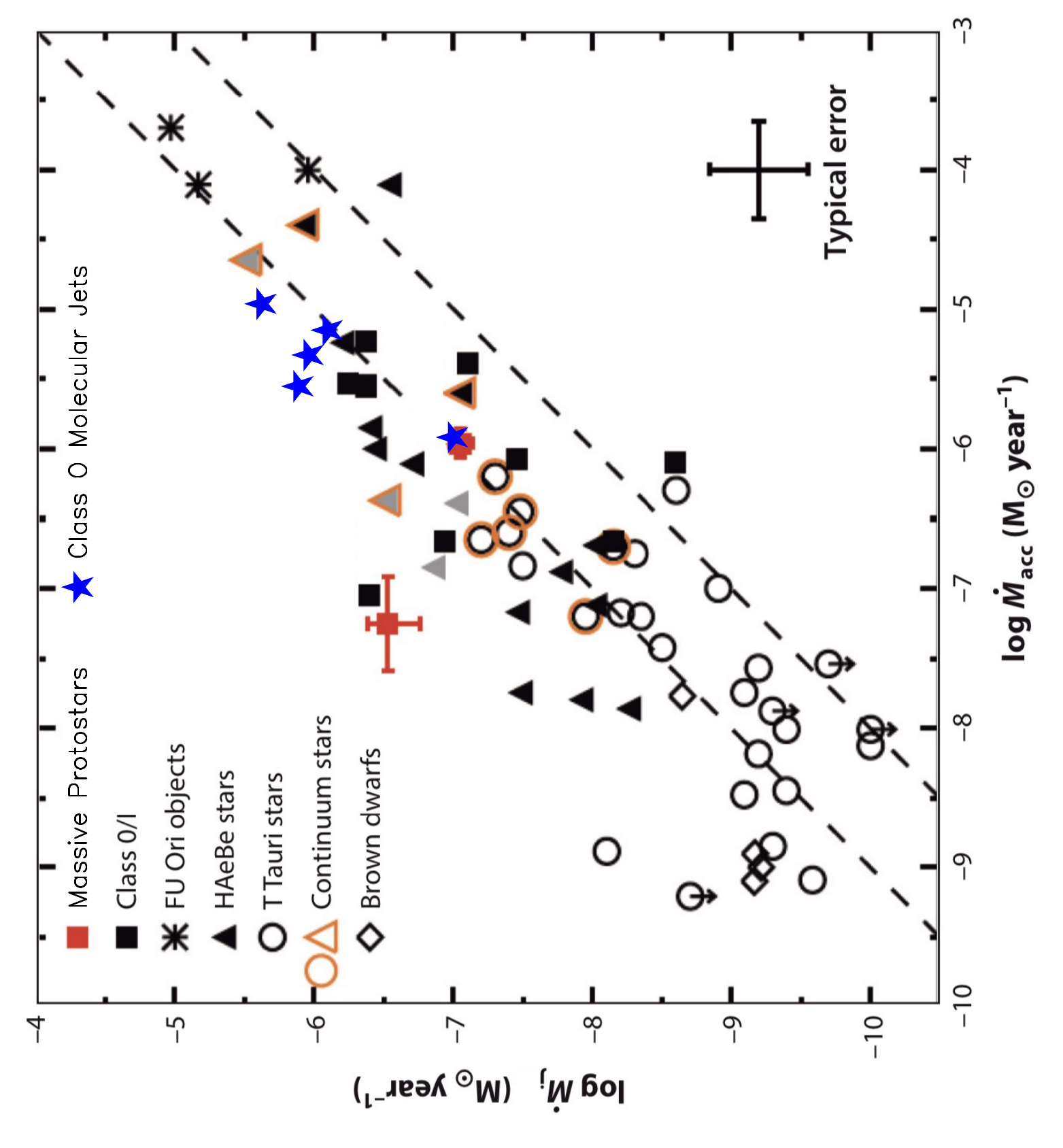}
\caption{Observed mass-loss rate in the jets $\dot{M}_{j}$ versus accretion rate
$\dot{M}_{\mathrm{acc}}$ of our jet sources (blue stars)
compared with those in different classes of YSOs associated with jets and outflows 
presented in \citet{Ellerbroek2013}.
The dashed lines indicate $\frac{\dot{M}_{j}}{\dot{M}_{\mathrm{acc}}}$ = 0.01 and 0.1.
\label{fig:MaccMjet}}
\end{figure}

The ratio of the mass-loss rate to the accretion rate has been used to probe
the mass-ejection efficiency that is then compared to the current
jet-launching models.  Figure \ref{fig:MaccMjet} shows the mass-loss rate
versus the accretion rate for the Class 0 jets here (as marked with blue
stars) in comparison to those found in different classes of young stellar
objects (YSOs) associated with jets and outflows presented in
\citet{Ellerbroek2013}.  As can be seen, our data points follow roughly the
trend found before, which shows the mass-loss rate and accretion rate
decreasing from the Class 0 to Class II (T-Tauri) phase from $10^{-6}$ to
$10^{-10} \solarmass \mathrm{\ yr}^{-1}$ and from $10^{-5}$ to $10^{-9}
\solarmass \mathrm{\ yr}^{-1}$, respectively.  Our data points here are in
the higher end with higher mass-loss rate and accretion rate, because the
jet sources here are younger.  A linear fit to our data points results in a
ratio $\frac{\dot{M}_{j}}{\dot{M}_{\mathrm{acc}}}\sim 0.19$, higher than
$\sim$ 0.12 obtained by fitting to all the Class 0/I data points.  However,
since the data distribution is rather scattered and our sample is small,
further work with more accurate measurements are needed to check this. 
Nonetheless, both ratios are consistent with magneto-centrifugal
jet-launching models.  Note that in current jet-launching models, there will
be wide-angle wind components around the jets, thus the mass-loss rate
estimated here could be a lower limit of the true value, and so could be the
ratio.


With the ratio, we can estimate the magnetic lever arm parameter
\citep{Shu2000,Pudritz2007}, which is defined as

\begin{equation}
\lambda \equiv \left(\frac{r_A}{r_0}\right)^2 \approx \frac{\dot{M}_{\mathrm{acc}}}{\dot{M}_j}
\end{equation}

\noindent where $r_0$ is the launching radius of the jets in the disks and
$r_A$ is the Alfv{\'e}n radius along the streamline launched from $r_0$. 
This parameter determines the extracted angular momentum and poloidal
acceleration, assuming a conservation of angular momentum and energy along
the streamline.  For example, the poloidal acceleration determines the
terminal velocity the jet can achieve.  Let $v_{\mathrm{kep}}$ be the
Keplerian velocity at the launching point, then $v_j \sim
\sqrt{2\lambda-3}\; v_{\mathrm{kep}}$, according to the current
magneto-centrifugal jet-launching models \citep{Shu2000,Pudritz2007}.  Thus,
with a given protostellar mass and jet velocity, we can derive the launching
radius with

\begin{equation} 
r_0 \sim (2\lambda-3)\; \frac{GM_\ast }{v_j^2}
\label{eq:ro_l}
\end{equation} 

\noindent Based on Table~\ref{tab:jetsource} using the 6 sources with both
mass and jet velocity measured, the jet sources have a mean mass of $M_\ast
\sim 0.10 \solarmass$ and the jets have a mean velocity of $v_j \sim$ 130
\vkm{}.  Then with a mean $\lambda \sim 1/0.19 \sim 5$ for our jet sources,
the mean jet launching radius is $\sim$ 0.04 au, which is $\sim$ $4 \, R_\ast$. 
Further observations of a larger sample with better measurements of all
related quantities are needed to refine this.



\section{Jet rotation} \label{sec:jetrotation}

In current jet-launching models, the jets are launched by
magneto-centrifugal force from the innermost parts of the disks, and are
thus expected to be rotating and magnetized.  In this way, angular momentum
can be carried away by the jets from the innermost parts of the disks,
allowing material there to actually fall onto the central protostars. 
Therefore, measuring jet rotation is an important task to confirm these
models and the role of the jets in removing the angular momentum from the
disks.  The measured amount of angular momentum in the jets can also help us
differentiate between the two competing models of jet launching, namely, the
X-wind model \citep{Shu2000} and the disk-wind model \citep{Konigl2000},
without spatially resolving the launching zones on the $\sim$ 0.05 au scale. 
These two models predict different amounts of angular momentum to be carried
away by the jets.  In particular, the jets in the X-wind model are launched
from the innermost edge of the disks and thus carry only a small amount of
specific angular momentum of $\lesssim 10$ au \vkm{} \citep{Shu2000}, while
the jets in the disk-wind model are launched from a range of radii from the
innermost edge out to $\sim$ 1 au, and thus can carry a larger amount of
specific angular momentum up to a few 100 au \vkm{} \citep{Pudritz2007}.


 Previously, jet rotation was tentatively detected in T-Tauri jets in
optical [OI], [NII], and [SII] lines using the STIS instrument aboard the
{\it Hubble Space Telescope}, with a velocity sampling of $\sim$ 25 \vkm{}
pixel$^{-1}$ and a spatial sampling of $\sim$ 7 au (or \arcsa{0}{05} in
Taurus) pixel$^{-1}$ \citep{Bacciotti2002,Coffey2007}.  The jets were found
to have a specific angular momentum up to a few 100 au \vkm{}, suggesting a
launching radius of 0.2 to 1 au \citep{Coffey2007}.  Similar measurements
were also made towards a few Class 0 jets, e.g., HH 212 \cite[with a spatial
resolution of $\sim$ 140 au and a velocity resolution of $\sim$ 1
\vkm{},][]{Lee2008}, HH 211 \cite[with a spatial resolution of $\sim$ 77 au
and a velocity resolution of $\sim$ 0.35 \vkm{},][]{Lee2009}, and NGC 1333
IRAS 4A \cite[with a spatial resolution of $\sim$ 480 au and a velocity
resolution of $\sim$ 0.67 \vkm{},][]{Choi2011} in SiO at radio wavelengths,
finding a launching radius of 0.03 to 2 au.  Tentative jet rotation was also
reported in the Class I jet HH 26 \cite[with a spatial sampling of $\sim$ 60
au and a velocity resolution of $\sim$ 34 \vkm{},][]{Chrysostomou2008} in
H$_2$, suggesting a launching radius of 2 to 4 au.  Tentative jet rotation
was also reported in SO in the intermediate-mass source Ori-S6
\citep{Zapata2010}, suggesting a launching radius of $\sim$ 50 au.  However,
all of these measurements are based on shock emission (e.g., SiO, SO, H$_2$,
and [OI]) and thus could be uncertain if the shock structures and
kinematics of the jets are not spatially resolved, for example, the
rotation in T-Tauri jets was later found to be false \citep{Coffey2012}.  In
addition, asymmetric shock structure and jet precession can also produce a
velocity gradient mimicking a jet rotation.

Since the jets are highly collimated and narrow, a spatial resolution of
better than 10 au is needed to spatially resolve them.  In addition, since
the rotation speed is small in the jets, a velocity resolution of $\sim$ 1
km/s is also needed to detect it.  More importantly, we need to zoom in to
the innermost parts of the jets, where the (internal) shocks have not yet
developed significantly, where the jets are not yet interacting
significantly with the surrounding and cavity material, and where the jet
precession effect is not significant, so that the rotation signature can be
preserved and the velocity gradient across the jet axis can be attributed to
jet rotation.  Moreover, since the jet radius decreases towards the central
source as discussed later, we also need to avoid zooming in too close to the
central source where the jets become unresolvable with current instruments.

\begin{figure}[htb]
\centering
\includegraphics[angle=270, width=\textwidth]
       {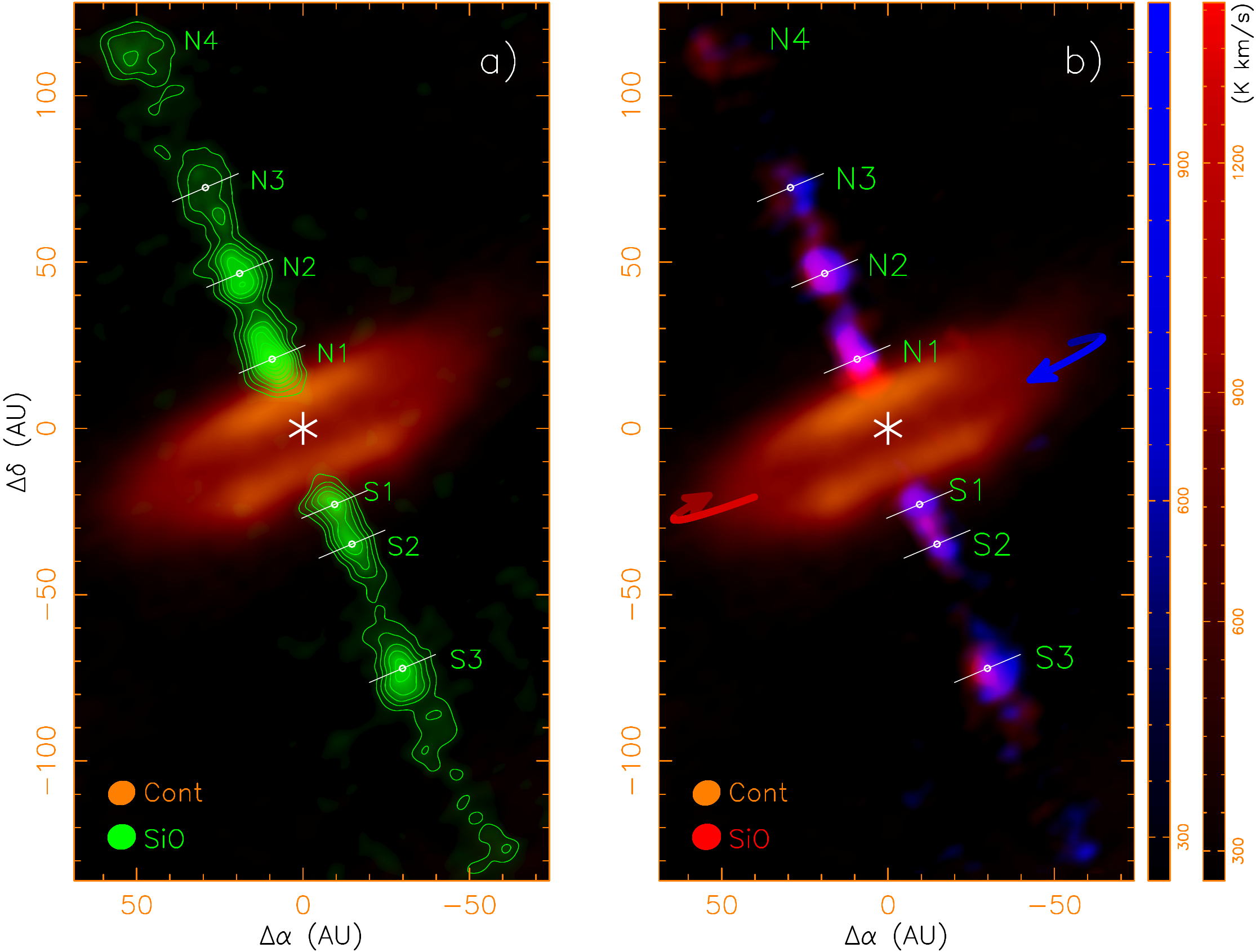} 
\caption{ALMA results of the SiO jet in HH 212 within 100 au of the central source, adopted
from \citet{Lee2017Jet}.
The orange image shows the dusty disk observed at 850 \micron{} \citep{Lee2017Disk}.
In (a), the green image shows the SiO jet. In (b), blueshifted emission and
redshifted emission are plotted separately to show the jet rotation around the jet axis.
The blue and red arrows show the disk rotation.
\label{fig:jetrot}}
\end{figure}

\begin{figure}[htb]
\centering
\includegraphics[angle=270, width=\textwidth]
       {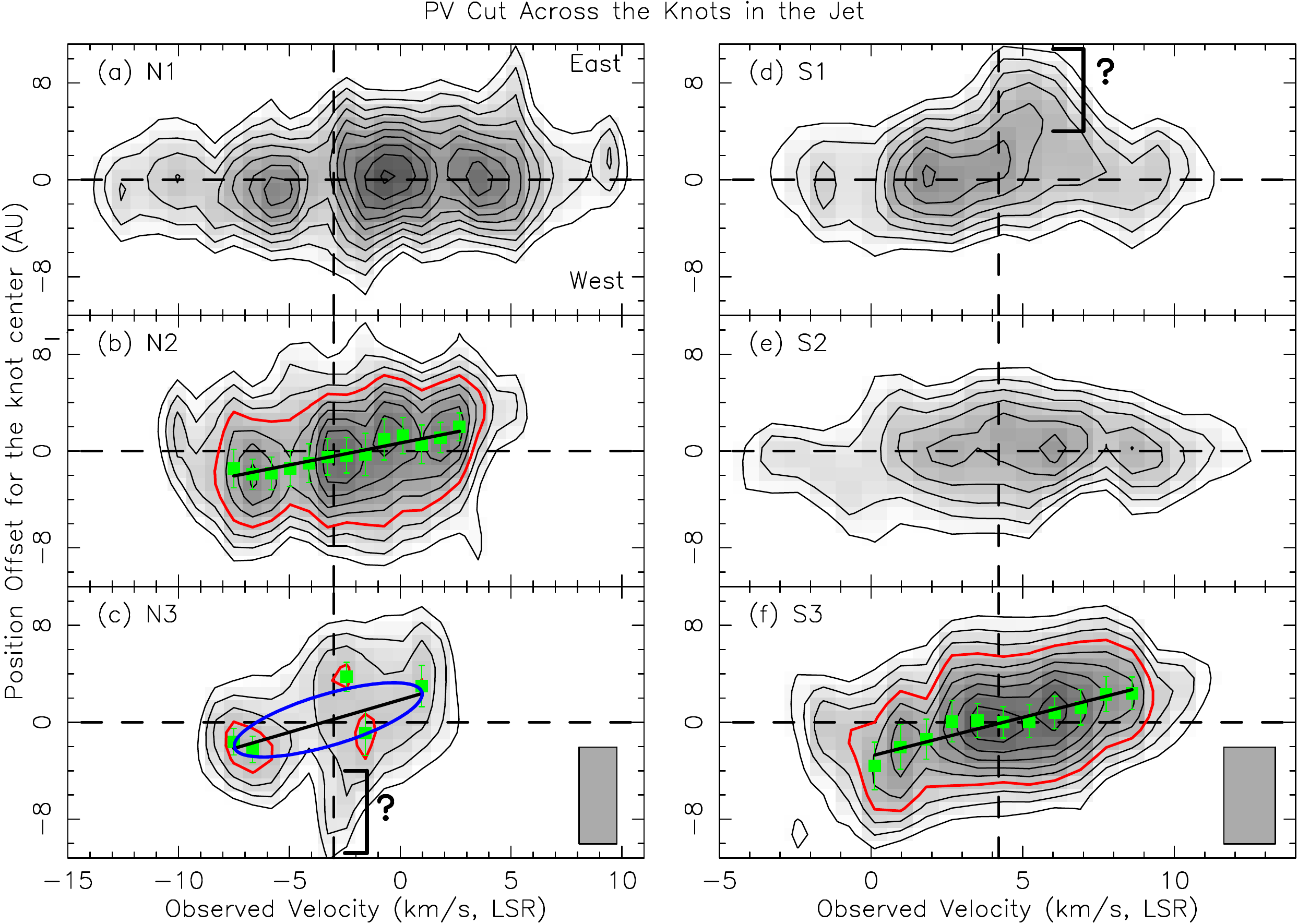} 
\caption{Position-velocity (PV) diagrams cut across the knots (N1-N3 and S1-S3) in the HH 212 jet,
adopted from \citet{Lee2017Jet}.
The horizontal dashed lines indicate the peak (central)
position of the knots. The vertical dashed lines indicate roughly 
the systemic (mean) velocities for the northern and southern jet components.
The red contours mark the 7$\sigma$ detections in knots N2, N3 and S3.
The solid lines mark the linear velocity structures across the knots
due to the jet rotation.
In (c), the blue ellipse is a tilted elliptical PV
structure expected for a rotating and expanding ring (see text).
\label{fig:jetpvHH212}}
\end{figure}

Thanks to the powerful Atacama Large Millimeter/submillimeter Array (ALMA)
with an unprecedented combination of high spatial resolution of $\sim$ 8 au
and high velocity resolution of $\sim$ 1 \vkm{}, \citet{Lee2017Jet} reported
a more reliable detection of jet rotation in the Class 0 jet HH 212 in SiO. 
This jet is almost in the plane of the sky, allowing us to measure the jet
rotation without being affected by projection effects.  In this jet, six
knots (N1, N2, N3 and S1, S2, S3) were detected in SiO within 100 au of the
central source, with three on each side, as shown in Figure
\ref{fig:jetrot}a.  As shown in Fig.~\ref{fig:jetrot}b, the blueshifted
emission and redshifted emission are on the opposite sides of the jet axis,
showing a consistent velocity gradient with the same velocity sense as the
disk rotation, indicating that the velocity gradient is due to jet rotation. 
Note that although the inner knots are less resolved, they still show a hint
of velocity gradient with the same velocity sense as the outer knots.  This
measurement strongly supports the role of the jet in carrying away the
angular momentum from the disk.  Based on the position-velocity diagrams
across the knots in the jet (Fig.~\ref{fig:jetpvHH212}), the specific
angular momentum in the knots is estimated to be $\sim$ 10 au \vkm{}, but
could be smaller because the knots are not well resolved spatially.


In the framework of magneto-centrifugal jet-launching models, the jet is the
central core of a magnetized wind.  The launching radius at the foot point
of the jet in the accretion disk can be derived from the observed specific
angular momentum and the observed velocity of the jet at a large distance
from the central source, assuming a conservation of energy and angular
momentum along the field line.  In addition, the wind can be assumed to have
enough energy to climb out of the potential well of the central star easily,
so that the kinetic energy of the wind is substantially greater than the
gravitational binding energy at the launching surface.  With these
assumptions, \citet{Anderson2003} has derived a useful relation between the
angular momentum at large distance and the jet launching radius ($r_0$) at
the foot point of the field line in the disk:


\begin{equation}
r_0 \approx 0.7 \, \textrm{au} \;
\left(\frac{l_j} {100\, \textrm{au} \;\vkme}\right)^{2/3}
\left(\frac{v_j} {100\, \textrm{km s$^{-1}$}}\right)^{-4/3}
\left(\frac{M_\ast} {1\,M_\odot}\right)^{1/3}
\label{eq:Anderson}
\end{equation}

\noindent where $l_j$ and $v_j$ are the specific angular momentum and
velocity of the jet at large distance.  In the case of HH 212 where $l_j$ is
small, we need to add two correction terms to the above solution as in the following
\citep{Lee2017Jet}

\begin{equation}
r_0 \approx 0.7 \, \textrm{au} \;
\left(\frac{l_j} {100\, \textrm{au} \ \vkme}\right)^{2/3}
\left(\frac{v_j} {100\, \textrm{km s$^{-1}$}}\right)^{-4/3}
\left(\frac{M_\ast} {1\, M_\odot}\right)^{1/3} \Big[1-\frac{2}{3}\eta + \frac{1}{9}
\eta^2\Big] \label{eq:jradius} 
\end{equation}

\noindent
with

\begin{equation} 
\eta = \frac{3}{2^{2/3}}(\frac{GM_\ast}{v_j l_j})^{2/3} \approx
0.38 \;\left(\frac{l_j} {100\, \textrm{au} \ \vkme}\right)^{-2/3}
\left(\frac{v_j} {100\, \vkme}\right)^{-2/3}
\left(\frac{M_\ast} {1\, M_\odot}\right)^{2/3} \,.
\end{equation}

\begin{figure}[htb]
\centering
\includegraphics[width=\textwidth]
       {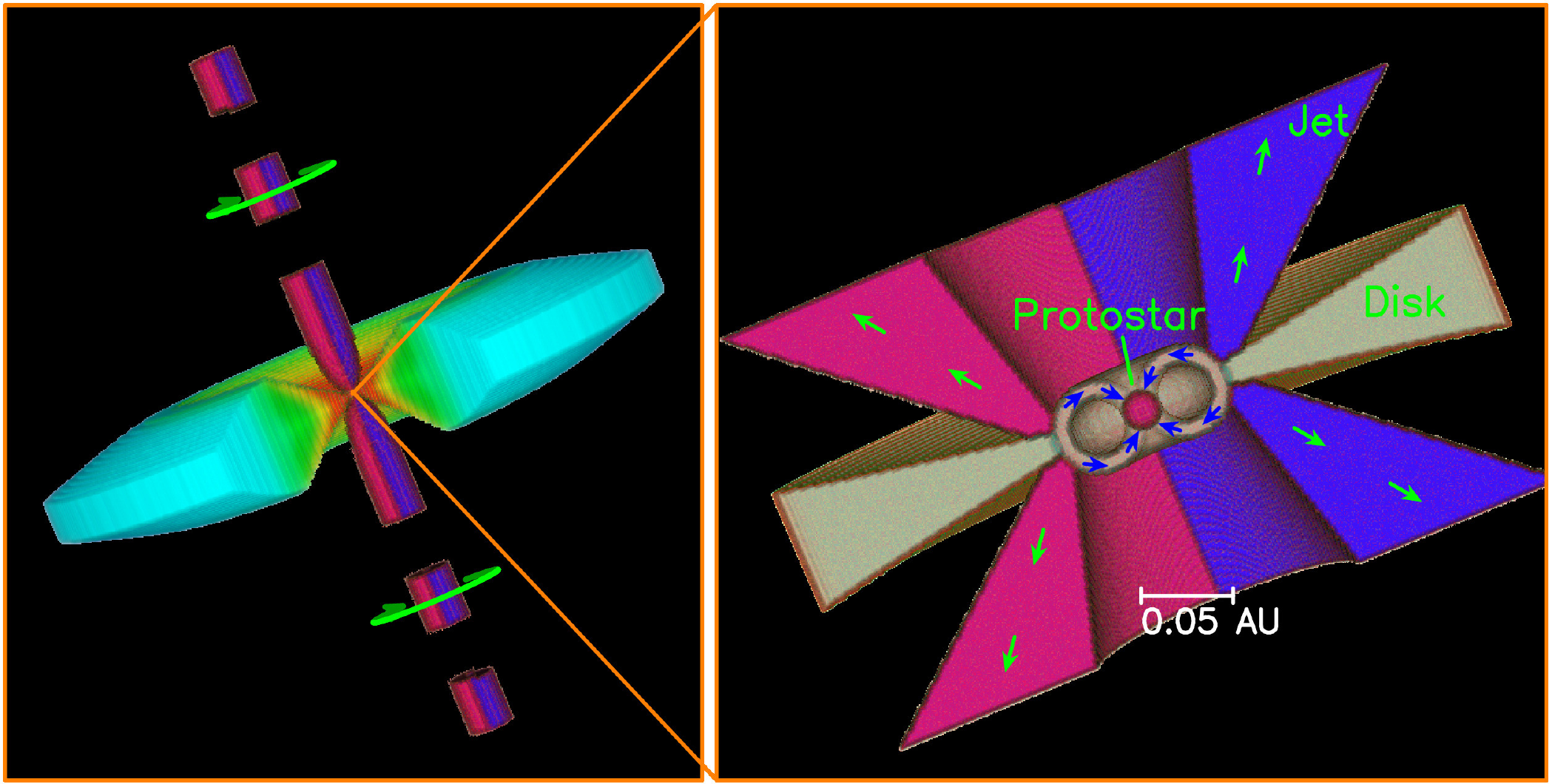} 
\caption{Cartoon showing a possible jet launching scenario for HH 212.
In the innermost part of the disk, part of the
disk material is ejected, forming a bipolar jet carrying the angular momentum
away, allowing the disk material to fall onto the central protostar through a
funnel flow.
\label{fig:jmodHH212}}
\end{figure}

The two correction terms can improve the accuracy of the launching
radius estimate in the case where the dimensionless parameter $\eta$ is not
much smaller than unity, particularly when the specific angular momentum
$l_j$ is relatively small, as is true for HH 212.  Then with $l_j \sim 10 $
au \vkm{}, $v_j \sim$ 135 \vkm{}, and $M_\ast \sim 0.25 \solarmass$ (see
Table \ref{tab:jetsource}), the launching radius of the HH 212 jet is
estimated to be $\sim$ 0.04 au, consistent with current models of jet
launching.  This results in a magnetic lever arm parameter of $\lambda \sim
3.2$ (see Eq.  \ref{eq:ro_l}), indicating that the jet has a specific
angular momentum about 3 times that in the disk at the launching radius. 
Since the mass-loss rate in the jet is estimated to be $\sim
1.3\times10^{-6} \solarmass$ (see Table~\ref{tab:jetsource}), the angular
momentum flux would be $1.3\times10^{-5} \solarmass$ au \vkm{} yr$^{-1}$. 
Fig.~\ref{fig:jmodHH212} shows a possible scenario of the jet launching in
HH 212 from the accretion disk.  In the innermost part of the disk, part of
the disk material is ejected, forming a bipolar jet carrying the angular
momentum away, allowing the disk material to fall onto the central protostar
through a funnel flow \citep{Shu2000}.

\begin{figure}[htb]
\centering
\includegraphics[width=0.9\textwidth]
       {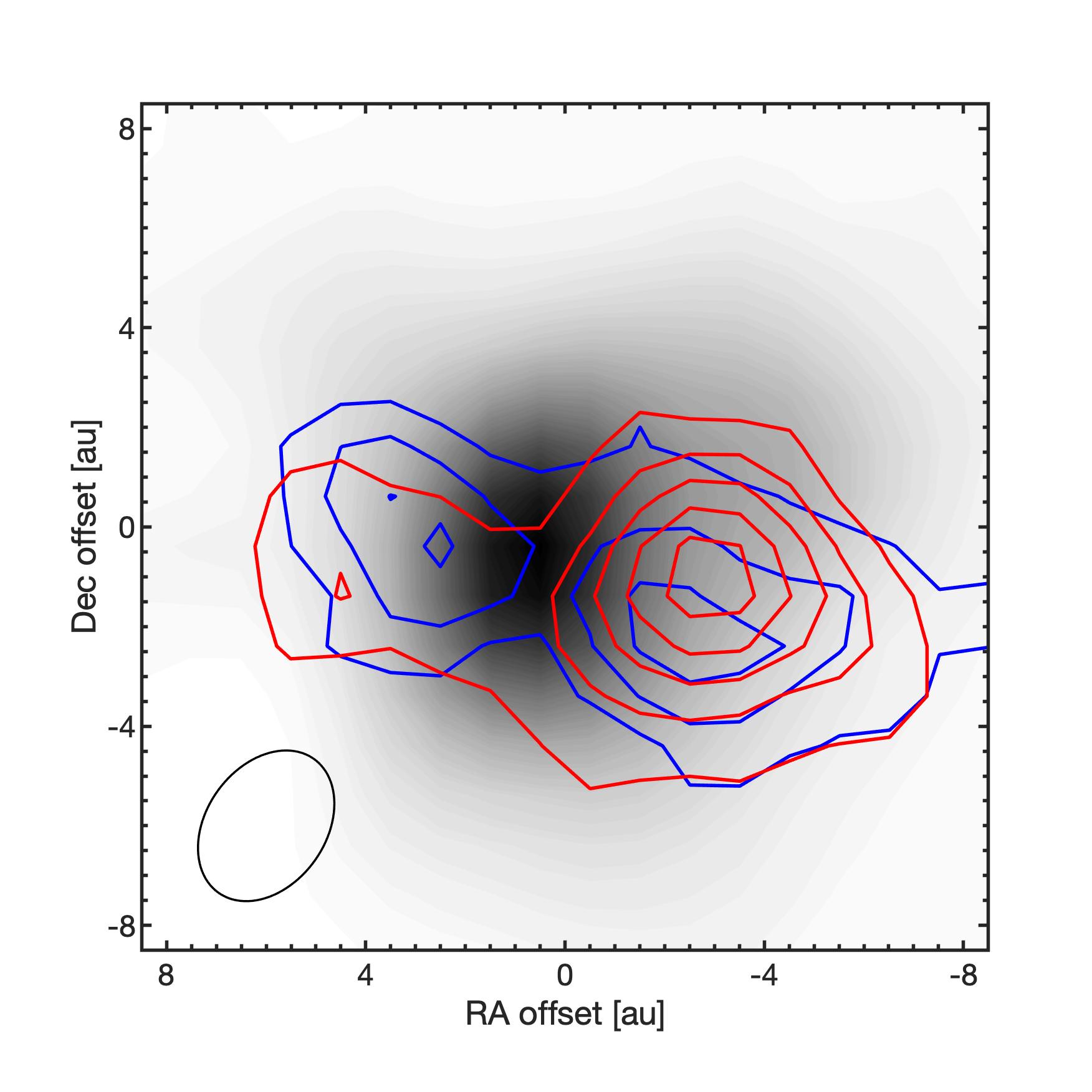} 
\caption{SiO maps of the jet in B335 obtained with ALMA at $\sim$ 3 au resolution
\citep{Bjerkeli2019}.
Blue and red contours show the blueshifted and redshifted emission maps 
of the jet in SiO J=5-4.
\label{fig:BjerkeliB335}}
\end{figure}

Recent ALMA observations at an unprecedented high resolution of $\sim$ 3 au
have detected a small SiO jet in B335 extending out from the central source
along the outflow axis \cite[][see also
Fig.~\ref{fig:BjerkeliB335}]{Bjerkeli2019}.  This jet is located within
$\sim$ 4 au of the central source and has a radius of $\lesssim$ 1 au.  No
clear rotation is detected in this jet, likely because the jet is not
resolved at current resolution.  In addition,  since the central source
seems to have a smaller protostellar mass of only $\sim 0.05 \solarmass$
\citep{Bjerkeli2019}, the jet rotation is expected to be smaller than that
seen in HH 212 and is thus more difficult to detect.  The SiO line in this
jet has a linewidth of $\sim$ 13 \vkm{} \citep{Bjerkeli2019}.  Assuming that
this linewidth is mainly due to jet rotation, then the specific angular
momentum of the jet is $\lesssim$ 6.5 au \vkm{}.  If this is the case, then
the jet launching radius would be $\lesssim$ 0.02 au, assuming a jet
velocity of 160 \vkm{} (see Table~\ref{tab:jetsource}).  Further
observations are needed to check this.

\section{Jet radius near the central source: collimation} \label{sec:jetradius}

\begin{figure}[htb]
\centering
\includegraphics[angle=270, width=\textwidth]
       {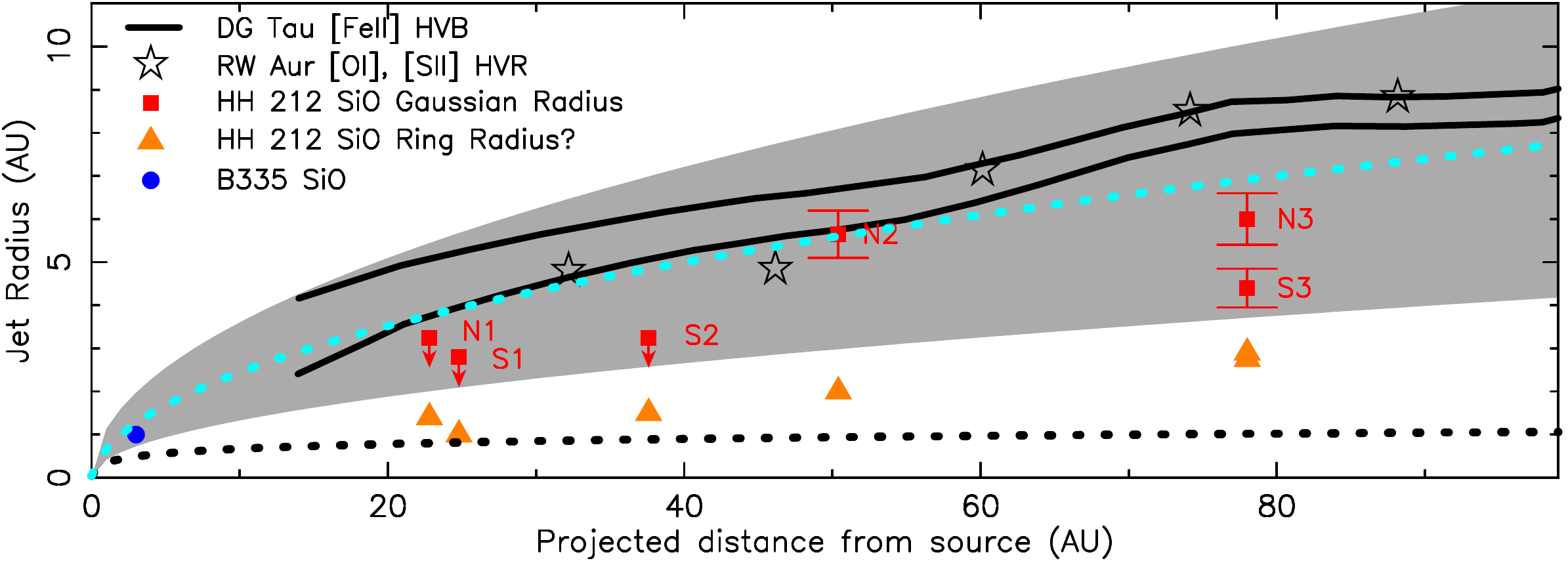}
\caption{Observed jet radius versus the distance from the central source.
Here, DG Tau data points are adopted from \citet{Agra-Amboage2011},
RW Aur from \citet{Woitas2002}, HH 212
from \citet{Lee2017Jet}, and B335 from 
\citet{Bjerkeli2019}. Note that in HH 212, 
the radius of knot N3 in HH 212 has
been revised to be $\sim$ 6 au here after excluding the emission in the west 
around $-3$ \vkm{} (see Fig.~\ref{fig:jetpvHH212}c, marked with a ``?''),
which could result from an interaction with the material in the cavity.
The gray shaded area is to guide the readers that the jet radius could be
increasing roughly parabolically with the distance.
The black dotted curve outlines the radius of a hollow cone 
in an X-wind launched at a radius of 0.04 au.
The cyan dotted curve indicates the streamline within which the X-wind 
contains 30\% of the total mass-loss rate.
The orange triangles indicate the radius of the SiO knots in HH 212, assuming
the SiO emission comes from a ring (see text).
\label{fig:jetradius}}
\end{figure}

With the unprecedented angular resolutions of ALMA, we can now measure the
radius of molecular jets within 100 au of the central sources in order to
probe the collimation zones, as done before for T-Tauri jets in the optical. 
Figure \ref{fig:jetradius} shows the radii of the jets in the two Class 0
sources HH 212 \cite[assumed to be half of the Gaussian widths reported
in][]{Lee2017Jet} and B335 \cite[measured from the SiO maps shown in 
Fig.~\ref{fig:BjerkeliB335} adopted from][]{Bjerkeli2019}
in comparison to those in the two T-Tauri sources, RW Aur \citep{Woitas2002}
and DG Tau \citep{Agra-Amboage2011}, previously measured from the
high-velocity emission maps in the optical.  As seen from the figure, the
Class 0 jets seem to be narrower and have a smaller expansion rate in jet
radius with distance than the T-Tauri jets.  However, further work with
a larger sample is needed to confirm it.  Interestingly, for both types of
jets, the jet radius could be increasing roughly parabolically with the
distance (i.e., with the square of the distance), as guided by the gray
shaded area in the figure, and thus roughly consistent with that predicted
in current magneto-centrifugal models of jet launching
\citep{Shu2000,Konigl2000}.  In these models, the jets are collimated
internally by their own toroidal magnetic fields and expand roughly
parabolically with the distance.  If the jet radius indeed expands roughly
parabolically with the distance, then the jets must have launched from the
innermost parts of the disks within a radius much less than 1 au of the
central sources, as guided by the gray shaded area.

\section{Are the jets hollow?}

In current magneto-centrifugal jet-launching models, there is a narrow
hollow cone in the jet center along the jet axis because of an intrinsic
expansion of the magneto-centrifugal winds coming from the disks.  This
hollow cone is also supported by an axially opened stellar magnetic field. 
However, there will be no hollow cone if the jets turn out to be stellar
winds launched by stellar magnetic field.  Therefore, it is critical to
check for the existence of a hollow cone in the jets in order to determine
the jet-launching models.

Here we can compare the observed radius of the HH 212 molecular jet to the
predicted radius of the hollow cone in the X-wind model \citep{Shu2000}. 
Since the launching radius of the jet is estimated to be $\sim$ 0.04 au, we
calculate the predicted radius of the hollow cone for this launching radius,
as shown as the black dotted curve in Fig.~\ref{fig:jetradius}.  For
comparison, we also plot the streamline (the cyan dotted curve) within which
the X-wind contains 30\% of the total mass-loss rate.  As can be seen, the
radius of the knots of HH 212 is larger than that of the hollow cone,
indicating that we can not rule out the existence of a hollow cone in this
jet.  In addition, the radius of the knots is smaller than the cyan
streamline, supporting that the jet traces the inner core of the wind.


We can also check the existence of the hollow cone by investigating the
kinematics.  The outer SiO knots (i.e., N2, S3, and N3) are roughly
resolved.  The position-velocity (PV) diagrams across them show a roughly
linear PV structure (see Fig.~\ref{fig:jetpvHH212}), suggesting that they
could actually be rotating rings, as expected if there is a hollow cone in
the jet.  For knot N3, the PV diagram even shows a tilted elliptical PV
structure (blue ellipse in Fig.~\ref{fig:jetpvHH212}c), as expected for a
rotating ring with an expansion \citep{Lee2018Dwind}.  Although the inner
knots (i.e., N1, S1, and S2) are not resolved, knots N1 and S2 also seem to
show the similar linear PV structure, suggesting that they could also be
rotating rings.  As a result, the kinematics across the knots in the HH 212
jet is consistent with a presence of a hollow cone in the jet.  If this is
the case, the radius of the knot is better assumed to be the radius of the
ring rather than the Gaussian radius.  Assuming that the SiO emission of each knot
comes from a ring, we can estimate the ring radius using the two ends of the
linear velocity structures in the PV diagrams.  As can be seen, the ring
radius of the knots (marked as orange triangles) is slightly larger than
that of the hollow cone, supporting that the SiO knots come from the
innermost core of the wind and the jet is hollow.


Since knot N3 shows an expansion in the PV diagram, we can also check if its
expansion velocity can be consistent with the intrinsic expansion velocity
of a magneto-centrifugal wind.  A rough fit (blue ellipse in Figure
\ref{fig:jetpvHH212}c) to the PV structure suggests an expansion velocity of
$\sim$ 3 \vkm{} in this knot.  Since the jet has a velocity of $\sim$ 135
\vkm{}, this expansion velocity would require a streamline with a very small angle of
$\sim$ 1.3\degree{} to the jet axis, and thus consistent with a streamline
in the innermost core of the wind around the possible hollow cone.







\section{Shock formation and sideways ejection}


Studying the formation of the knots and bow shocks in the molecular jets
allows us not only to further constrain the origins of the jets but also to
understand the jet chemistry.  Since the molecular jets, e.g., HH 211
\citep{Gueth1999,Lee2010HH211}, HH 212 \citep{Lee2008,Lee2015}, and L1448 C
\citep{Hirano2010}, appear to be continuous in the inner parts, the jet
ejection at the base should be continuous.  Some of them show roughly equal
spacings in between knots and bow shocks, indicating that the knots and bow
shocks are produced by quasi-periodical variations in ejections.  Since the
knots and bow shocks are well detected in, e.g., SiO and H$_2$, they trace
strong shocks, as in the optical jets \citep{Ray2007,Hartigan2011,Bally2016}.
Therefore, they can not merely trace quasi-periodical
ejections of enhanced jet densities, which alone would not produce shocks
\citep{Frank2014}.  There must also be a quasi-periodical variation in
ejection velocity, so that a shock can be formed as the fast jet material
catches up with the slow jet material.  A simple form of velocity variation
is a sinusoidal variation, presumably induced by an orbital motion of the
perturbations.  An eccentric orbit can even cause an enhanced accretion rate
and thus mass-ejection rate near periastrons \citep{Reipurth2000,Benisty2013}.


Many hydrodynamical simulations have been performed to study the formation
of knots and bow shocks with a periodical sinusoidal variation of ejection
velocity \cite[e.g.,][]{Raga1990,Stone1993,Suttner1997,Lee2001}.  In the
body of the jet, the fast material catches up with the slower material,
forming an internal shock (which consists of a forward shock, a backward
shock, and an internal working surface in between).  This internal shock is
first seen as a knot.  As it propagates down along the jet axis, because of
sideways ejection in the shock, it expands laterally and grows to a wider
knot and then a bow shock and then an internal shell of post-shock gas
closing back to the central source \citep{Lee2001}.  Each bow shock will
show a spur velocity structure in the PV diagram along the jet axis
\citep{Lee2001,Ostriker2001}, because the gas velocity decreases rapidly
away from the bow tip.  If the amplitude of the velocity variation is
larger, the resulting bow shock is bigger.

A periodical variation in jet velocity has been suggested in the jet in
L1448 C \citep{Hirano2010}.  In particular, each SiO knot has its higher
velocity in the upstream (closer to the jet source) side and lower velocity
in the downstream side.  The opposite velocity gradient is always seen in
the faint emission between the knots.  Such a velocity pattern is likely to
be formed by a periodical variation in the ejection velocity, as seen in the
simulations \citep{Stone1993,Suttner1997}.  In HH 211 (a jet close to the
plane of the sky), the innermost pair of jetlike structures BK1 and RK1,
which consists of a chain of smaller subknots, show a velocity range
decreasing with distance \citep{Lee2010HH211}, also as expected for the
knots being formed by a periodical variation in velocity
\citep{Suttner1997}.  The two knots further away are believed to be newly
formed internal shocks, each showing a pair of backward and forward shocks
expanding longitudinally with time \citep{Jhan2016}.  In these jets, a
velocity variation of 20--30 \vkm{} has been suggested to produce the knots
that have a spacing corresponding to a period of a few ten yrs
\citep{Hirano2010,Jhan2016}.

\begin{figure}[htb]
\centering
\includegraphics[width=0.4\textwidth]
       {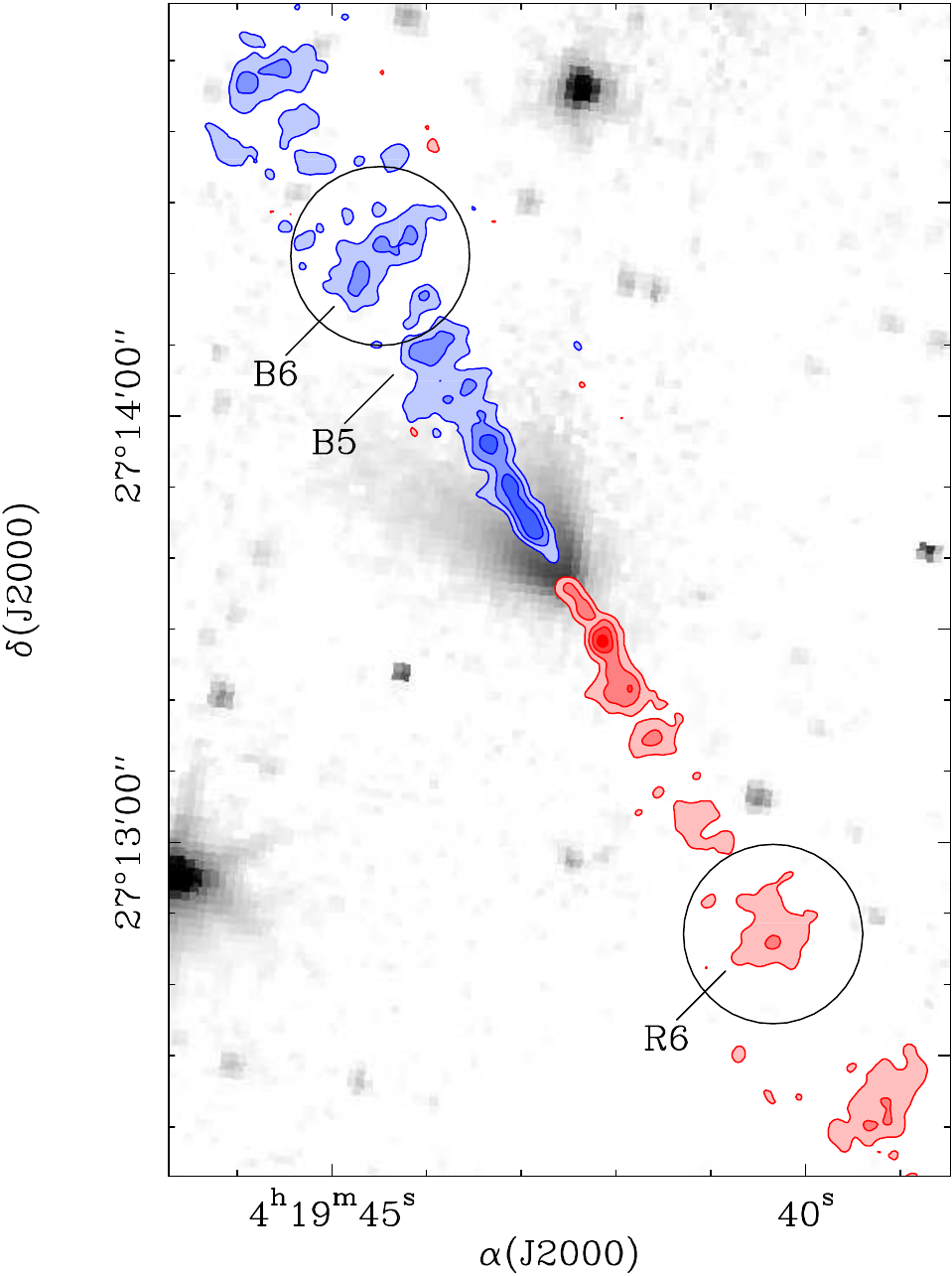}
\includegraphics[width=0.4\textwidth]
       {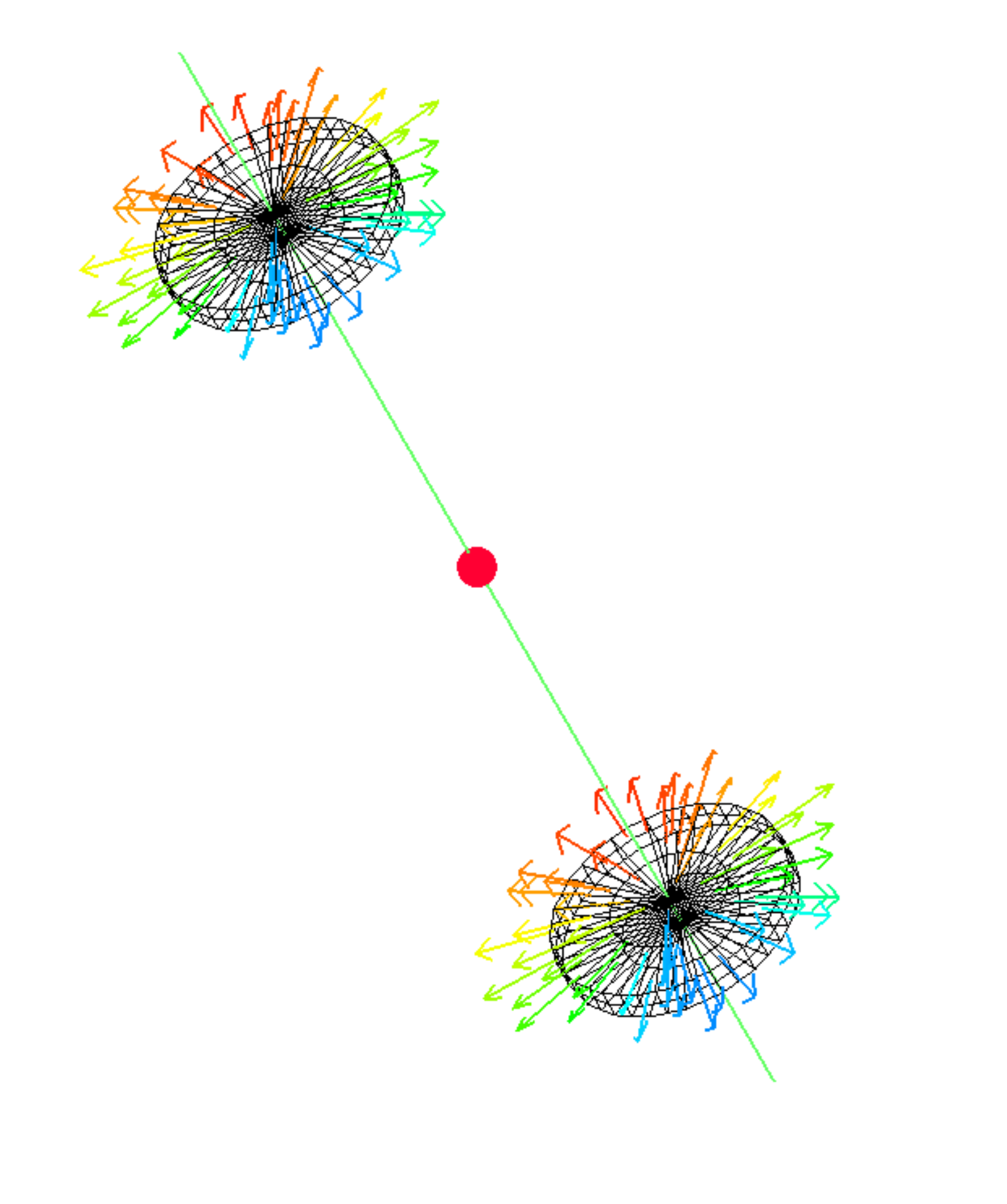}
\caption{Molecular jet in IRAS 04166+2706:
(Left) A chain of knots are detected in CO along the jet axis \citep{Santiago2009}. 
(Right) A schematic diagram showing the sideways ejection in
knots B6 and R6, as concluded from 
a detailed ALMA kinematic study \citep{Tafalla2017}.
\label{fig:sideways}}
\end{figure}

A clear example showing the sideways ejection of the internal shocks is the
molecular jet in IRAS 04166+2706.  It consists of at least 7 pairs of knots
seen in SiO and CO, with the width increasing with distance from the central
source \cite[][see also Fig.~\ref{fig:sideways} left]{Santiago2009}.  The
velocity field along the jet axis shows a sawtooth pattern for each knot. 
This pattern, together with a systematic widening of the knots with distance
to the central source, is consistent with them tracing the
laterally-expanding internal shocks viewed at a high inclination angle to
the plane of the sky \citep{Stone1993,Suttner1997}.  At a large distance
from the central source, the knots, B6 and R6, have grown to possess
bow-like structures.  A detailed kinematic study of this pair of knots with
ALMA confirmed that they are internal bow shocks where material is being
ejected laterally away from the jet axis \citep{Tafalla2017}, as shown in
Fig.~\ref{fig:sideways}.

\begin{figure}[htb]
\centering
\includegraphics[width=0.6\textwidth]
       {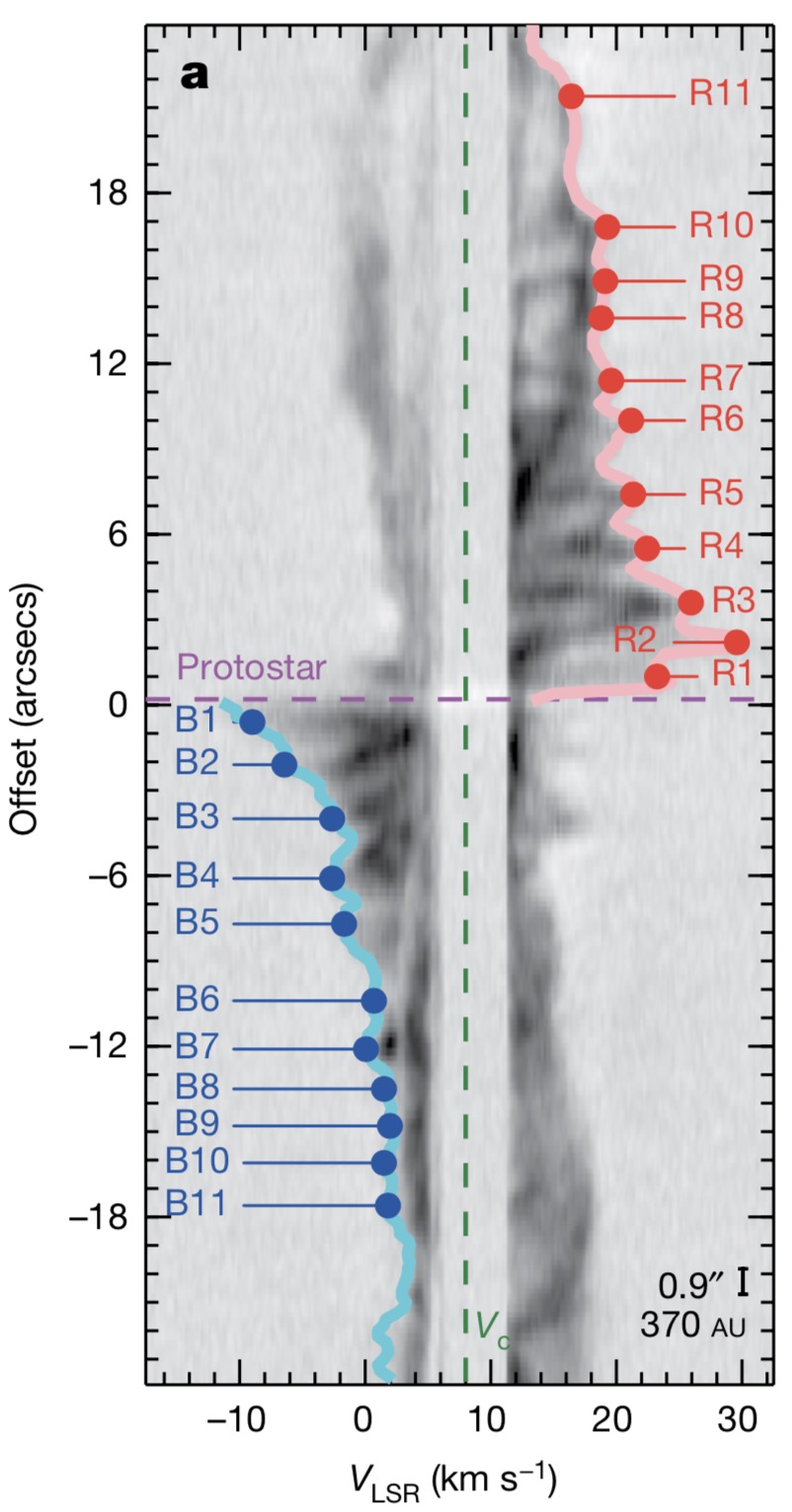}
\caption{Position-velocity diagram of the CO jet in CARMA-7 along the jet axis \citep{Plunkett2015}.
Labels B1 to B11 and R1 to R11 mark the positions of the internal knots and bow shocks.
\label{fig:pvC7}}
\end{figure}

Sideways ejection is also detected in other jets.  In HH 212, the sideways
ejection is clearly seen in the two spatially resolved internal bow-like
knots SK4 and SK5 in CO and SiO \citep{Lee2015}.  Since the HH 212 jet is
almost in the plane of the sky, the knots are associated with arclike
velocity patterns \citep{Stone1993}, instead of a sawtooth velocity pattern
seen in a highly inclined jet.  Nested internal shells are also
seen in CO, with each extending from a H$_2$ bow shock back to the central
source \citep{Lee2015}, as seen in the simulations.  Sideways ejection is
also detected in CARMA-7.  In this source, a chain of small CO bow shocks is
seen along the jet axis, with each bow shock associated with a spur velocity
structure in the PV diagram along the jet axis \cite[][see also
  Fig.~\ref{fig:pvC7}]{Plunkett2015}, consistent with laterally-expanding
internal bow shocks.  The spur structure is slightly tilted because of a
small inclination of the jet to the plane of the sky.


\section{Periodical variations and their possible origins}

As discussed earlier, some of the molecular jets show roughly equal spacings
in between knots and bow shocks that can be due to quasi-periodical
variations in ejections, allowing us to probe quasi-periodical perturbations
of the underlying accretion process in the disks.  In addition, the rotating
disks associated with some molecular jets are now resolved and found to have
a radius smaller than 50 au (see Table~\ref{tab:jetsource}), providing a
strong constraint on the possible perturbations in the disks.  Therefore, we
can now further constrain the origins of the quasi-periodical variations in
ejections.

\begin{figure}[htb]
\centering
\includegraphics[angle=180, width=\textwidth]
       {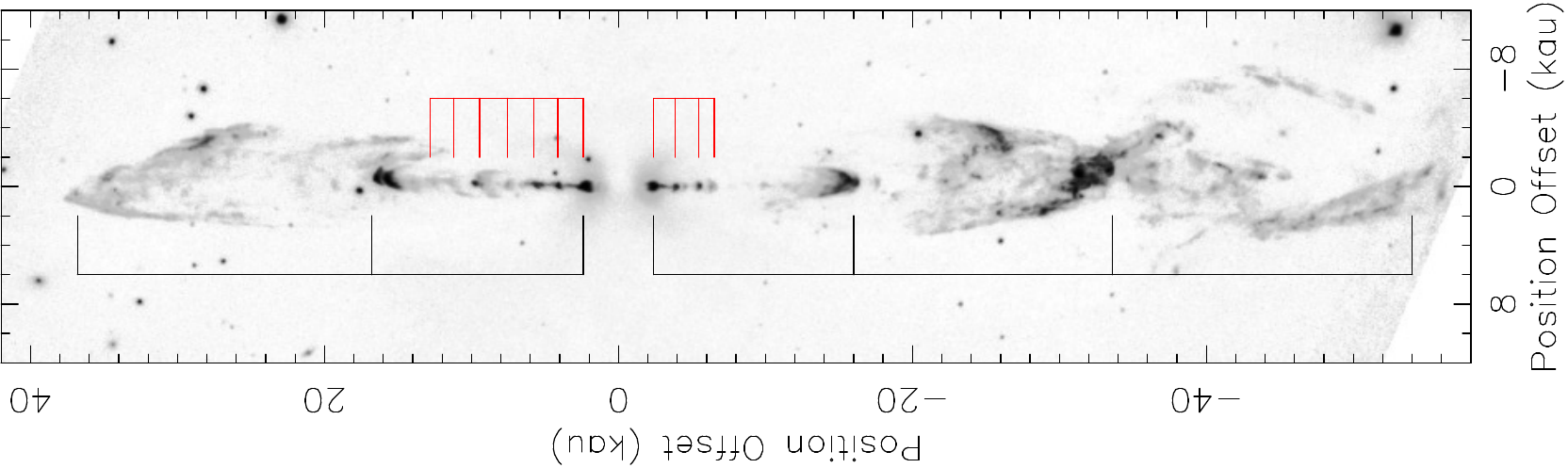}
\caption{Spacings between knots and major bow shocks in HH 212 seen in H$_2$ \citep{Zinnecker1998}.
Red lines show the spacing for the inner knots and black lines for the prominent bow shocks.
\label{fig:spacingHH212}}
\end{figure}

The period of variation can be obtained by dividing the knot or bow shock
spacing by the jet velocity.  Previously in the optical jets in the later
phase, at least two periods of variations, one short and one long, have been
found to operate at the same time in one single jet.  For example,
\citet{Raga2002} found two periods of 270 yr and 1400 yr in HH 34, and two
periods of 60 yr and 950 yr in HH 111.  Similarly in the molecular jet HH
212 in the early phase, \citet{Zinnecker1998} also found two periods, one
for the inner knots with a spacing of $\sim$ 1700 au and the other for the
prominent bow shocks with a spacing of $\sim$ 17200 au (see Figure
\ref{fig:spacingHH212}).  Adopting a mean jet velocity of $\sim$ 135 \vkm{}
(see Table 1), these spacings correspond to periods of 60 yrs and 605 yrs,
respectively.  In recent high-resolution ALMA observations, an additional
knot spacing of $\sim$ 30 au (see Fig.~\ref{fig:jetradius}) with a
corresponding period of $\sim$ 1 yr is also detected near the jet source. 
Such a short period of variation was also detected before in, e.g., the DG
Tau jet which has a period of $\sim$ 2.5 yr \citep{Agra-Amboage2011}.  It is
believed that such a short period of variation will be found in more jets
near the central sources when more high-resolution observations are
obtained.  


\begin{table} 
\small 
\centering
\caption{Periodical variations in well-defined molecular jets}
\label{tab:jetperiod} 
\begin{tabular}{lllll} 
\hline Source & Spacing & Period & a &  References  \\
              & (au) & ($yr$) & (au) &  \\ \hline\hline
IRAS 04166+2706 & 1065 & 83 & ? &  1 \\ 
                & 5330 & 415 & ? &  1 \\ 
NGC1333 IRAS4A2 & 1600 & 76 & 9 & 2 \\ 
L1157 & 15000  & 660 & 26 &  3 \\
HH 211  & 256   & 12  & 2   &   4 \\  
           & 800   & 38  & 5   &   4 \\  
           & 4300  & 204 & 15  & 4 \\ 
L1448 C& 680    & 20   &  3   & 5 \\ 
           & 8000   & 240  &  17  & 5 \\ 
HH 212  & 30    & 1   & 0.6 & 6 \\ 
           & 1700  & 60  & 10  &  6 \\ 
           & 17200 & 605  & 45  & 6 \\ 
\hline
\multicolumn{5}{p{11cm}}{
Here $a$ is the corresponding orbital radius of the perturbation defined by Eq. \ref{eq:binsep}.
References: 
(1) \citet{Santiago2009}, 
(2) \citet{Choi2011}
(3) \citet{Kwon2015} 
(4) \citet{McCaughrean1994}, \citet{Lee2010HH211},
(5) \citet{Hirano2010} 
(6) \citet{Zinnecker1998}, \citet{Lee2017Jet}
}   \\
\end{tabular}
\end{table}

\begin{figure}[htb]
\centering
\includegraphics[angle=270, width=\textwidth]
       {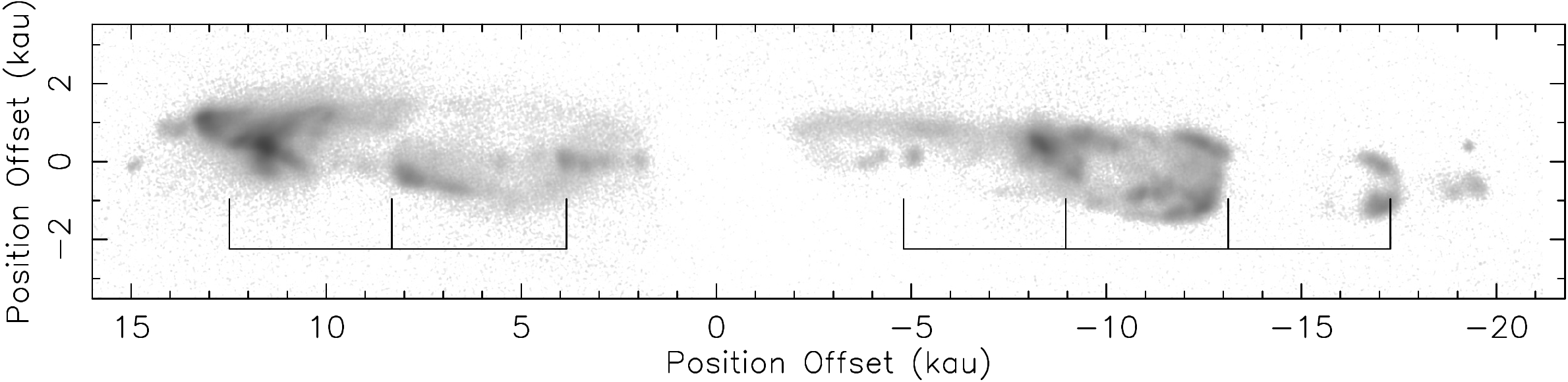}
\caption{Spacing between major bow shocks in HH 211 seen in H$_2$ \citep{Hirano2010}.
\label{fig:spacingHH211}}
\end{figure}

Table \ref{tab:jetperiod} lists the periods of variations estimated from six
well-defined molecular jets.  These periods range from 1 yr to 660 yrs and
can be broadly divided into 3 different groups: periods of a few yrs
associated with knots near the jet sources, periods of a few ten yrs
associated with knots and small bow shocks in the inner parts of the jets,
and periods of a few hundred yrs associated with prominent bow shocks (see
Fig.~\ref{fig:spacingHH212} for HH 212 and Fig.~\ref{fig:spacingHH211}
for HH 211).  Recent ALMA observations of a jet in the Class 0 protostar
CARMA-7 show a clear chain of small CO bow shocks along the jet axis, with a
spacing of $\sim$ 680 au \citep{Plunkett2015}, or a period of $\sim$ 32 yrs
assuming a velocity of 100 \vkm{}, and thus belongs to the period group of a
few ten yrs.  These periods could come from periodical perturbations of the
underlying accretion in the disks on the orbital time scales.  In order to
study this possibility, we derive the corresponding orbital radii for these
periods with the following equation from the Kepler's third law of orbital motion:

\begin{equation} 
a = \left(\frac{G M_\ast T^2}{4 \pi^2}\right)^{1/3} = \left(\frac{M_\ast}{M_\odot}\right)^{1/3}
 \left(\frac{T}{\textrm {yr}}\right)^{2/3} \, \textrm{au} 
\label{eq:binsep}
\end{equation}

\noindent where $T$ is the period and $M_\ast$ is mass of the central protostar.



As can be seen from Table \ref{tab:jetperiod}, periods of a few hundred yrs
correspond to orbital radii of a few tens of au.  Interestingly in HH 212
and HH 211, the implied orbital radii of the perturbations are similar to
the radii of their rotating disks, which have a radius of $\sim$ 44 and 16
au (see Table \ref{tab:jetsource}), respectively.  For other jet sources,
their disks are not resolved yet and further observations are needed to
check this possibility.  Therefore, it seems that the longest period of
perturbations in the disks can come from the outer edge of the disk,
probably induced by gravitational instability (GI) powered by envelope
accretion.  Such instability has been detected in the embedded Class I disk
in HH 111, which shows a pair of spiral arms extending from the outer edge
of the disk to the inner disk where the Toomre Q parameter is of the order
of unity \citep{Lee2019HH111}.  An accretion shock is also detected around
that disk in SO, indicative of an active accretion from the envelope
\citep{Lee2016HH111}.  In simulations, protostellar disks could be
periodically gravitationally unstable, forming spiral arms in the disks and
thus producing enhanced accretion rates onto the protostars periodically
\citep{Tomida2017}.  These periodical enhanced accretion rates can then
cause periodical major ejections (such as FU Ori-like bursts) of the jets,
producing the prominent bow shocks in the jets.  A recent study of a
few FU Ori-type objects have detected arms and fragmented structures that
can be attributed to gravitationally unstable disks \citep{Takami2018}, also
supporting this possibility.  In this scenario, older protostars, which
have larger masses and disks, would have periods of a few thousand yrs.  For
example, HH 111, which has a protostellar mass of $\sim 1.5 \solarmass$ and
a disk radius of $\sim$ 160 au \citep{Lee2019HH111}, would have a period of
$\sim$ 1650 yrs.  HH 111 has a length of more than 10 pc
\citep{Reipurth2001}, and deeper observations in H$_2$ are needed to check
for this period.

The next group of periods has a mean of $\sim$ 50 yr, with a mean
corresponding orbital radius of $\sim$ 6 au.  This radius could be the
location where the deadzone is located with negligible ionization.  One
possible disturbance source is a close binary companion, which could evolve
from a secondary fragmentation in a second core collapse
\citep{Machida2008}.  The possibility of a binary companion was also
suggested in, e.g., HH 212 \citep{Lee2007HH212}, L1448 C \citep{Hirano2010},
and HH 211 \citep{Lee2010HH211} because of the wiggle in their jets.  Other
origins could be stellar magnetic cycles or global magnetospheric
relaxations of the star-disk system \citep{Frank2014}.  Alternatively,
perhaps GI can penetrate to the inner parts of the disks, forming dense
(protoplanetary) clumps in spiral arms, causing enhanced accretion rates onto
the protostars \citep{Vorobyov2005}.

The shortest period is $\sim$ 1 yr in HH 212, corresponding to an orbital
radius of $\sim$ 0.6 au.  Similarly in the DG Tau jet, which is driven by a
T-Tauri star with a mass of $\sim 0.7 \solarmass$, the corresponding
orbital radius for its 2.5 yr period is $\sim$ 1.5 au
\citep{Agra-Amboage2011}.  It is not clear what origin can introduce this
perturbation on this small orbital scale.  It could be due to a periodical
turn on of magneto-rotational instability (MRI) \citep{Balbus2006} in the
innermost parts of the disks where the temperature is high (1000 K) and the
ionization is sufficient \citep{Audard2014}.  It could also be related to
the phase transition from dust-rich to dust-free region in the disks,
because the jets are likely launched from dust-free regions in the innermost
parts of the disks, as discussed later.













\section{Wiggles and binaries?}

Two types of wiggles are detected in protostellar jets in the early phase,
as in the later phase.  One is point-symmetric (i.e., S-shaped) wiggles that
could be due to precession of the jets \cite[e.g.,][]{Raga1993}, which in
turn could be due to precession of the accretion disks because of tidal
interactions in noncoplanar binary systems \cite[see, e.g.,][]{Terquem1999}. 
The other is reflection-symmetric (i.e., C-shaped or W-type) wiggles
\citep{Fendt1998,Masciadri2002,Lee2010HH211,Moraghan2016} that could be due to orbital
motion of the jet sources around a binary companion.  Both types of wiggles
can take place simultaneously in single jets \citep{Raga2009}.
 Unlike most of the optical jets, molecular jets are seen on both sides,
allowing us to better determine the types of the wiggles and derive the
binary properties.  Combining with the total mass derived from other
methods, we can also derive the mass of each binary component.  In addition, the
periods of these wiggles could be compared to those of the ejection
variations in producing the knots and bow shocks in the jets, providing
further constraints on the potential binary formation in the central regions
of star formation in the early phase.  In particular, formation of a close
binary can start in the early phase before the collapse of a second core
and even during the protostellar phase because of angular momentum
redistribution at the center of the system \citep{Machida2008}.

\begin{figure}[htb]
\centering
\includegraphics[width=0.45\textwidth]
       {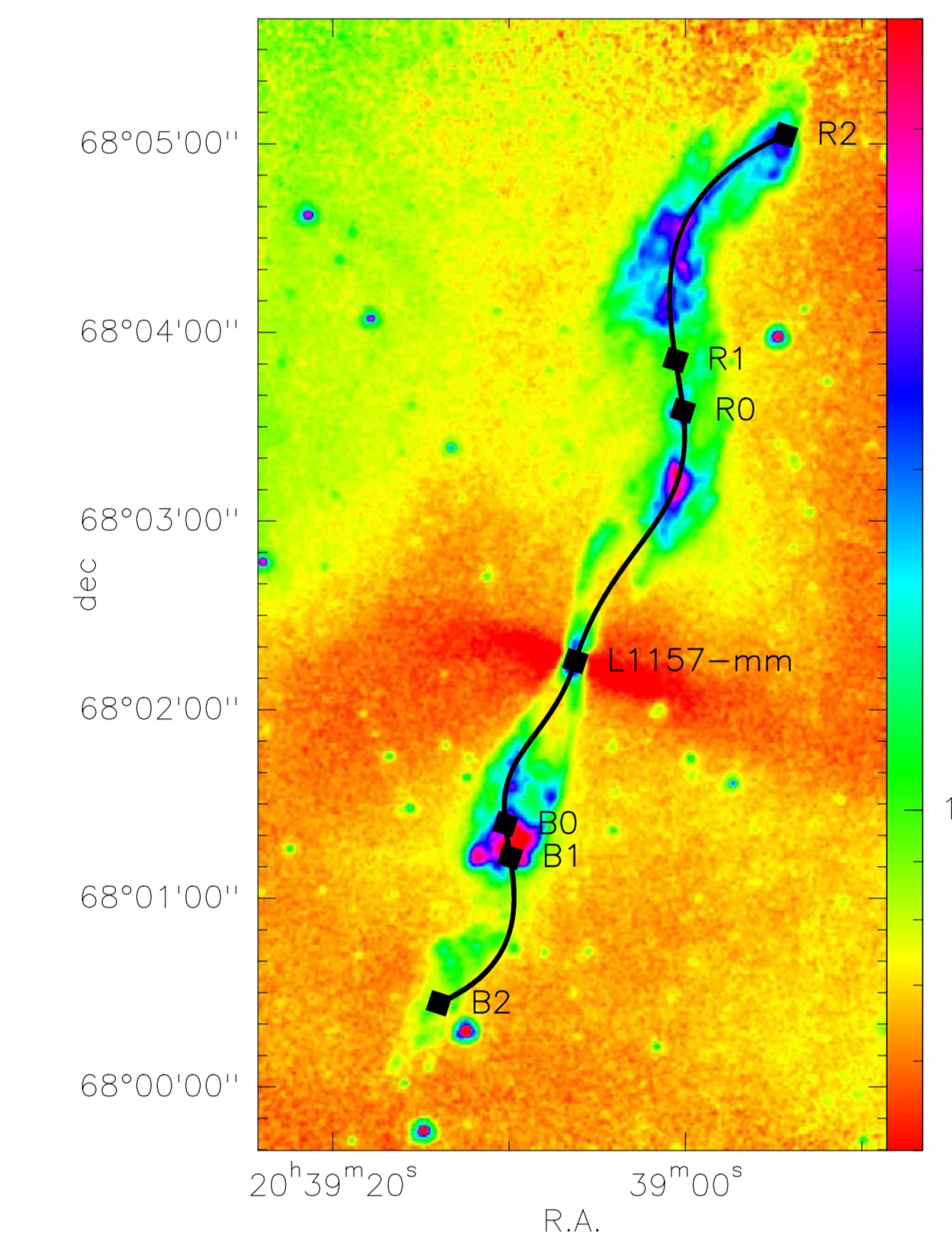}
\includegraphics[width=0.45\textwidth]
              {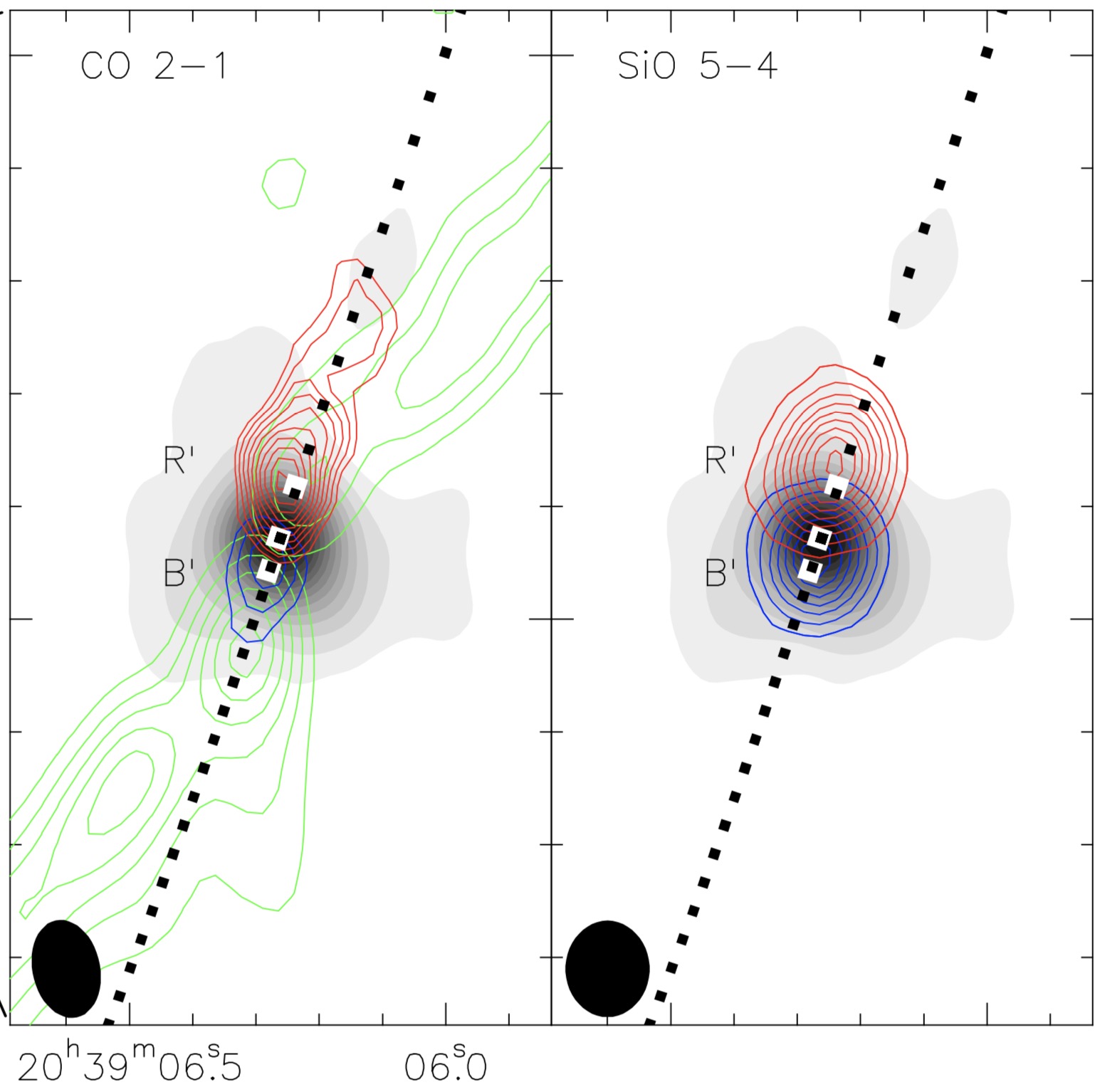}
\caption{L1157 jet and outflow adopted from \citet{Podio2016}. (Left)
A precessing jet model plotted on the Spitzer/IRAC 8 \micron{} map of the outflow.
(Middle) Blueshifted and redshifted CO emission of the jet.
(Right)  Blueshifted and redshifted SiO emission of the jet.
\label{fig:L1157Prec}}
\end{figure}

The young source L1157 has a S-shaped outflow and thus a precessing jet was
proposed to produce its outflow morphology \cite[][see also Figure
\ref{fig:L1157Prec} Left]{Gueth1996,Bachiller2001,Takami2011}.  Recently,
\citet{Kwon2015} proposed that there could be two precessing jets, each
along one side of the outflow cavity walls, to produce the outflow
morphology.  However, \citet{Podio2016} detected only a single jet in SiO
and CO within 200 au of the central source propagating along the symmetry
axis of the outflow lobes (see Fig.~\ref{fig:L1157Prec} Middle and Right),
confirming the previously proposed one precessing jet model
\citep{Gueth1996,Bachiller2001}.  The precession period is $\sim$ 1640 yrs,
about twice that estimated from the bow shock separation (see Table
\ref{tab:jetperiod}).

One mechanism to produce a precessing jet is disk precession induced by
tidal interaction with a binary companion in a non-coplanar orbit
\cite[e.g.,][]{Terquem1999}.  The jet source has a low-mass accretion disk
(with an outer radius $r_d$), which is precessing as a result of the
perturbations due to the companion.  Let the jet source have a mass $M_j$
and a companion a mass $M_c$ and the binary has a separation $a$ in a
circular orbit.  The ratio of precession period $\tau_p$ to orbital period
$\tau_o$ can then be given by

\begin{equation}
\frac{\tau_p}{\tau_o} 
=\frac{32}{15 \cos \alpha} \left(\frac{a}{r_d}\right)^{3/2} \left(1+\frac{M_c}{M_j}\right)^{1/2} \left(\frac{M_j}{M_c}\right)
\end{equation}

\noindent where $\alpha$ is the half-opening angle of the resulting
precession cone \citep{Terquem1999,Anglada2007,Raga2009}. 
\citet{Terquem1999} suggested that the tidal truncation of the accretion
disk will lead to a ratio $a/r_d=2-4$.  The jet precession in L1157 could be
due to this disk precession.  Taking a mean ratio of $a/r_d \sim 3$ and
assuming $M_C \sim M_J$, we have $\frac{\tau_p}{\tau_o} \sim 15$ and thus
the orbital period $\tau_o \sim$ 110 yrs.  With Equation \ref{eq:binsep},
the binary separation is estimated to be $\sim$ 8 au, and thus the disk
radius is $\sim$ 2.7 au.  Further work is needed to check this possibility. 

Another possible mechanism is an asymmetric envelope accretion onto the
disk.  Simulations have shown that a large misalignment between the magnetic
axis and rotation axis in the star-forming core can produce a flattened
infalling envelope and a rotating disk at the center.  However, the major
axis of the flattened infalling envelope will be misaligned with the major
axis of the disk \citep{Hirano2019}, as also seen in the observations of HH
211 \citep{Lee2019HH211}.  In this case, the envelope accretion onto the
disk will be asymmetric, causing the disk and the jet to precess
\citep{Hirano2019}.   Indeed, the jet axis in HH 211 has already been
found to have precessed by $\sim$ 3\degree{} in the past
\citep{Eisloffel2003}, supporting this scenario.


On the other hand, reflection-symmetric wiggle has been suggested in the jets
HH 211 \citep{Lee2010HH211,Moraghan2016} and HH 111 \citep{Noriega2011}, and could be due
to an orbital motion of the jet sources.  As discussed in
\citet{Masciadri2002} and \citet{Lee2010HH211}, we can derive the period
$P_o$, velocity $v_o$, and radius $R_o$ of the orbit of the jet source
around the companion from three measurable quantities: the jet velocity
$v_j$, the half-opening angle $\kappa$ and the periodic length (i.e.,
wavelength) $\Delta z$ of the wiggle, with the following equations:

\begin{equation}
P_o = \frac{\Delta z}{v_j} \;,\hspace{1cm} v_o \approx \kappa \; v_j \;,
\hspace{1cm}R_o \approx \frac{\kappa \Delta z}{2 \pi}
\end{equation}

\noindent Again, assuming the jet source has a mass $M_j$ and the
companion a mass $M_c$, and $M_j=m M_c$, then the binary separation
would be $a=(1+m)R_o$.
From Kepler's third law of orbital motion, 
the total mass of the binary would be

\begin{eqnarray} 
M_t \approx  9.5\times 10^{-4}\ (1+m)^3 
\times \left(\frac{\eqt{v}{j}}{100 \;\vkme{}}\right)^2 \left(\frac{\kappa}{1^\circ}\right)^3
\frac{\triangle z}{100 \, \mathrm{au}}\, M_\odot 
\end{eqnarray} 

\noindent In HH 211, with $\Delta z \sim 1750$ au, $\kappa \sim 0.0094$ (or
0.54\degree), $v_j \sim 110$ \vkm{}, and $M_t \sim 0.08 \solarmass$
\citep{Lee2019HH211}, we have $P_o \sim$ 83 yrs, $v_o \sim 0.94$ \vkm{},
$R_o \sim 2.6$ au, $m \sim 2.1$, and $a \sim 8.2$ au.  The mass of the jet
source would be $\sim 0.053 \solarmass$.  The orbital period is $\sim$ 2
times that generates the knot separation, which is $\sim$ 38 yrs (see Table
\ref{tab:jetperiod}).  
It is possible that the orbit is eccentric and two perturbations could
generate two knots per orbit.  This could occur either through the
periastron passage to the companion and a close approach to the inner edge
of the circumbinary disk, or two close approaches to the inner edge of the
circumbinary disk per eccentric orbit
\citep{Moraghan2016}. Since the binary separation here is
about half of the disk radius, observations at higher resolution are needed
to resolve the binary and check this possibility.  On the other hand, in HH
111, with $\Delta z \sim 86400$ au, $\kappa \sim 0.013$ (or 0.74\degree{}),
and $v_j\sim$ 240 \vkm{} \citep{Noriega2011}, and a total mass of $\sim 1.5
\solarmass$ \citep{Lee2010HH111,Lee2016HH111}, we have $P_o \sim$ 1710 yrs,
$v_o \sim 3.1$ \vkm{}, and $R_o \sim 178$ au, $m\sim -0.1$, and $a \sim 163$
au.  Thus, within the uncertainty, the jet source has almost no mass,
inconsistent with it driving a powerful jet  longer than 10 pc
\citep{Reipurth2001}.  In addition, the binary separation would be similar
to the disk radius, which is $\sim$ 160 au \citep{Lee2019HH111}, but no
binary companion was detected there around the edge of the disk
\citep{Lee2019HH111}.  Thus, further work is needed to check the
reflection-symmetric wiggle in the HH 111 jet.

\section{Magnetic fields in the jets}

For high-mass protostars, synchrotron radiation have been detected in a few
protostellar jets, e.g., HH 80-81 \citep{Marti1993} and W3(OH)
\citep{Wilner1999}, at centimeter wavelengths, indicating a presence of
magnetic fields in the jets.  Magnetic field morphology can thus be derived
from the polarization pattern of the synchrotron radiation.  So far,
polarization has only been detected in HH 80-81 \citep{Carrasco2010}.  It was
detected at a spatial resolution of $\sim$ 20,000 au towards the jetlike
structures at a large ($\gtrsim$ 30,000 au) distance from the central
protostar, where the underlying jet interacts with the ambient material
\citep{Rodriguez-K2017}.  The implied magnetic fields there are mainly
poloidal, with a field strength of $\sim$ 0.2 mG.  There is also a hint of
toroidal fields near the edges of the jetlike structures.  However, more
sensitive observations are needed to detect the polarized emission and thus
the field morphology in the underlying jet itself, which is much narrower with
a radius of $\sim$ 2000 au \citep{Rodriguez-K2017}.  Recent observations in
W3(H$_2$O) \citep{Goddi2017} also detected linear and circular polarizations
in the water masers at the center of the synchrotron jet within tens to
hundreds of au from the central source.  They suggested that the magnetic
field could evolve from having a dominant component parallel to the outflow
velocity in the pre-shock gas, with field strengths of a few tens of mG, to
being mainly dominated by the perpendicular component of a few hundred of mG
in the post-shock gas where the H$_2$O masers are excited.



For low-mass protostars, the jets are less energetic and thus do not have
strong enough synchrotron radiation and water masers for polarization
detection with current instruments.  Fortunately, when they are young, they
have a high content of molecular gas because of high mass-loss rate
\citep{Glassgold1991,Shang2006,Cabrit2007,Lee2007HH212,Hirano2010}, allowing
us to map their magnetic fields using the linear polarization in molecular
lines with the so-called Goldreich--Kylafis (GK) effect
\citep{Goldreich1981,Goldreich1982}.  In the presence of a magnetic field, a
molecular rotational level splits into magnetic sublevels, producing a line
polarization with its orientation either parallel or perpendicular to the
magnetic field.  Recent successful detection of this GK effect in the SiO line
in HH 211 confirmed this method of mapping the field morphology in the jets
from the low-mass protostars \citep{Lee2018Bjet}.  In the future, sensitive
observations of the Zeeman effect in molecular lines will allow us to derive the
strength of the magnetic fields \citep{Cazzoli2017}.


\begin{figure}[htb]
\centering
\includegraphics[angle=270, width=\textwidth]
       {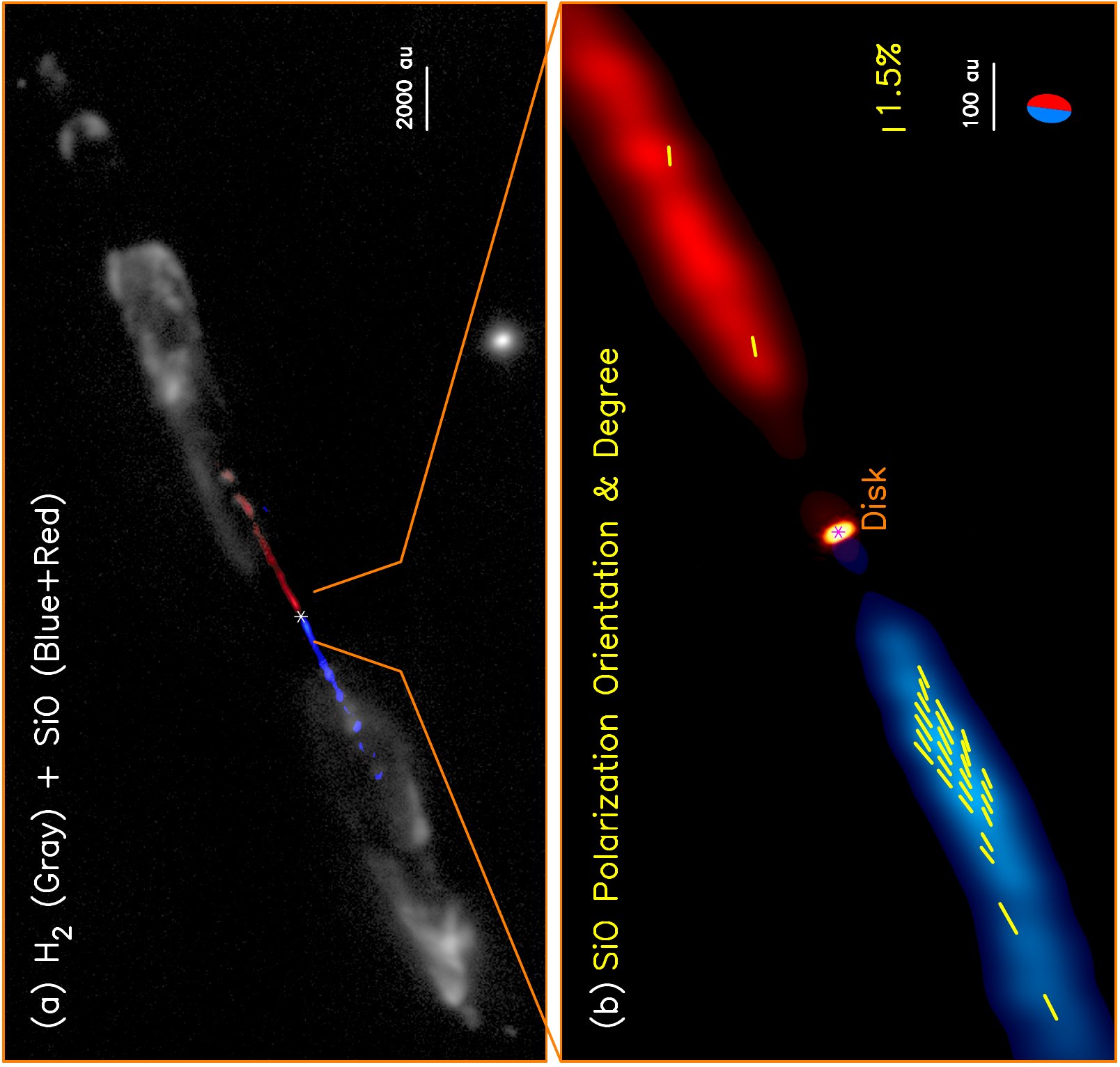}
\caption{SiO J=8-7 line polarization detection in the HH 211 jet \citep{Lee2018Bjet}.
The asterisk marks the position of the central driving source.
Gray image shows the outflow and the outer part of the jet in H$_2$ adopted 
from \citet{Hirano2006}. Blue image
and red image show the blueshifted and redshifted jet components in SiO, respectively. 
Yellow line segments indicate the polarization orientations seen in SiO.
\label{fig:Bjet}}
\end{figure}

Figure \ref{fig:Bjet} shows the SiO line polarization in J=8-7 transition
detected in the HH 211 jet within a few hundred au of the protostar.  The
jet is best seen in SiO in the J=8-7 transition.  At this transition, the
optical depth is close to 1 \citep{Lee2009} and the collision rate is lower
than the radiative transition rate for a typical jet density of
$10^{6}$--$10^{7}$ \cmc{}, both optimal for polarization from the GK effect
\citep{Goldreich1981,Goldreich1982}.  As can be seen, the polarization
orientations are all roughly parallel to the jet axis.  The implied magnetic
fields could be either mainly poloidal or mainly toroidal and additional
polarization observations in another SiO transition line are needed to
resolve this ambiguity in the field morphology, as done before in a CO outflow
\citep{Ching2016}.  It could be mainly toroidal, as suggested in current
jet-launching models, in order to collimate the jet at large distances.  The
field strength (projected on the plane of the sky) was estimated to be
$\sim$ 15 mG \citep{Lee2018Bjet}.  This field strength is about 2 times the
toroidal field strength expected from a typical X-wind model \citep{Shu1995}
for a low-mass protostellar jet like HH 211, which is reasonable considering
a shock compression in the jet and all the uncertainties in the
measurements.

\def\nH2{n_{\textrm{\scriptsize H}_2}}



The jets may have poloidal fields as well. In HH 212, the jet is found to be
wiggling, but with the amplitude of the wiggle being saturated at some
distance, inconsistent with a typical jet precession that has an amplitude
increasing linearly with the distance.  The saturation in the amplitude may
suggest a current-driven kink instability in the jet \citep{Cerqueira2001},
see, e.g., Fig.~3 in both \citet{Cerqueira2001} and \citet{Mizuno2014}. 
For the kink instability to take place, we have the Kruskal--Shafranov
criterion $|B_p/B_\phi| < \lambda/ 2 \pi A_m$ \cite[e.g.,][]{Bateman1978},
where $A_m$ is the maximum displacement and $\lambda$ is the wiggle
wavelength.  In HH 212, with $A_m \sim$ \arcsa{0}{1} and $\lambda \sim$
\arcsa{5}{6}, we have $|B_p/B_\phi| < 9$.  Therefore, this kink instability,
if in action, could be initiated in the central part of the jet, where the
magnetic field is dominated by the poloidal field \citep{Pudritz2012}.  This
poloidal field could serve as a ``backbone" to stabilize the jet
\citep{Ouyed2003}.  The toroidal field dominates only near the jet edges in
order to collimate the jet.



\section{Origin of molecular gas in the jets}

Unlike those in the Class I and Class II phases, the jets in the Class 0
phase appear to be mainly molecular and well detected in CO, SiO, and SO. 
Note that near the base of the jets, free-free emission from ionic gas can
also be detected in the Class 0 phase
\citep{Rodriguez1997,Reipurth2002,Reipurth2004,Rodriguez2014,Tobin2016Per}.  It is possible that the
high molecular content in the Class 0 phase is due to the high mass-loss
rate and thus the fast formation rate of the molecular gas.  Since the
mass-loss rate in the jets is high ($\sim 10^{-6} \solarmass$ yr$^{-1}$),
molecules such as CO, SiO, and SO could have formed via gas-phase
reactions in an initially atomic jet close to the launching point within 0.1
au \citep{Glassgold1991}.  The abundances of SiO and SO in the gas phase are
found to be highly enhanced in the jets as compared to those in the
quiescent molecular clouds, even close to within 30 au of the central
sources where the dynamical timescale is $<$ 1 yr
\citep{Lee2017Jet,Lee2018Dwind,Bjerkeli2019}.


As discussed earlier, based on the ejection efficiency, jet rotation, and
the expansion of jet radius near the jet sources, the molecular jets in the
Class 0 phase should be launched within $\sim$ 0.1 au of the jet sources. 
Most of these jet sources have a bolometric luminosity $\gtrsim 1
\solarlum$, suggesting that the dust sublimation radius in their accretion
disks is $\gtrsim$ 0.1 au \citep{Millan2007}, which is outside the inferred
jet launching radius.  In the case of HH 212, \citet{Tabone2017} proposed
that the SiO jet could be the innermost part of a disk wind launched at
$\sim$ 0.05--0.2 au, by modeling the PV structures of the SiO emission
across the jet axis (see their Figs.~2e and 2f).  Since their model PV
structures show emission peaks at the two high-velocity ends instead of at the
low velocity on the jet axis, the major part of the jet must still be
launched significantly closer than 0.2 au.  In any case, since the jet
source in HH 212 has a bolometric luminosity of $9 \solarlum$, the
sublimation radius should be $\sim$ 0.2 au \citep{Millan2007}, and thus
still larger than the launching radius.


It is likely that elements such as Si, S, C, and O, are already released
from the grains into the gas phase at the base of the jets.  Since the jets
are well collimated with a high mass-loss rate of $\sim 10^{-6}
\solarmass \mathrm{\ yr}^{-1}$ (see Table \ref{tab:jetsource}), SiO, CO, and SO can
form quickly within $\sim$ 0.1 au because of the high density in the jet
\citep{Glassgold1991}.  The jets are bright in SiO J=8-7, which has a high
critical density of $\gtrsim 10^7$ \cmc{}, further supporting the high density in
the jets.
In addition, the Si$^+$ recombination and SiO
formation are expected to be faster than the photodissociation caused by
possible far-ultraviolet radiation of the central protostar
\citep{Cabrit2012}.  Since these molecules are fully released from the dust
grains at the base, the observed abundances in the jets are higher
as compared to those in the quiescent molecular clouds.  


For the jet sources with a bolometric luminosity $< 1 \solarlum$, e.g.,
B335 and IRAS 04166+2706 (see Table \ref{tab:jetsource}), the dust
sublimation radius is $<$ 0.1 au \citep{Millan2007}.  However, it is still
possible that their jets are launched from a dust-free zone, because their
jet launching radius could be smaller, as discussed earlier.  In B335, since
the mass-loss rate is only $\sim$ $10^{-7} \solarmass \mathrm{\ yr}^{-1}$, further
work is needed to check if the density in the jet is high enough to form the
molecules in the gas phase.  In this source, the SiO J=5-4 emission is only
detected within a few au of the jet source, probably because the mass-loss
rate is low.


In case the jets are launched from dusty zones, SiO abundance in the jets
can be enhanced as a consequence of grain sputtering or grain-grain
collisions in the shocks releasing Si-bearing material into the gas phase,
which reacts rapidly with O-bearing species (e.g., O$_2$ and OH ) to form
SiO \citep{Schilke1997,Caselli1997}.  Another possible explanation is that
the SiO molecules existed on the grain mantles and are released into the gas
phase by means of shocks as suggested by \cite{Gusdorf2008}.  Similarly,
sulfur can be released from dust grains in the form of H$_2$S in the shocks
and that is then oxidized to SO \citep{Bachiller2001}.  Alternatively, the
SO molecules might be abundant on dust grains and directly released from
grains in the shocks \citep{Jimenez2005,Podio2015}.




\section{Disk winds around the jets?}


As discussed earlier, protostellar jets can carry angular momentum away from
the innermost region of the disks, allowing the disk material to feed the
central protostars.  However, other mechanisms are also needed to transfer
angular momentum outward within the disks or carry it away from the disks,
so that the disk material can be transported to the innermost region from
the outer region of the disks.  For example, MRI can turn on in the inner
part of the disks where the temperature is high ($\gtrsim$ 1000 K)
\citep{Audard2014} to transfer angular momentum outward.  GI can also be
excited in the outer part of the disks to transfer angular momentum outward,
as suggested in many simulations \citep{Bate1998,Tomida2017} and by the
detections of a pair of spirals in the embedded disk in HH 111
\citep{Lee2019HH111} and probably in the older disk in Elias 2-26
\citep{Perez2016}.  Besides, low-velocity extended tenuous disk winds can be
present as well to carry angular momentum away from the disks.  In
particular, previous observations have shown that wide-angle radial winds
are needed to drive molecular outflows
\citep{Lee2000,Lee2001,Hirano2010,Arce2013}, especially in the later phase
of star formation \citep{Lee2005L43}.


In the two popular magneto-centrifugal jet-launching models from the disks,
e.g., the X-wind model and the disk-wind model, the jets are merely the
central cores of the winds.  Therefore, in addition to the jets, the
models also predict wide-angle tenuous winds around the jets.  In the X-wind
model, the wide-angle wind comes from the same disk radius as the jet within
$\sim$ 0.05 au of the central source.  This model has been proposed to
produce the molecular outflows and the collimated jets simultaneously
\citep{Shang2006}.  On the other hand, in the disk-wind model, an extended
(wide-angle) disk wind comes from a range of radii up to $\sim$ 20 au
around the central jet, extracting angular momentum from the disks at larger
radii.  The detections of poorly collimated and low-velocity rotating
molecular outflows near the disks in a few older sources, including CB 26
(Class II) \citep{Launhardt2009}, TMC1A (Class I) \citep{Bjerkeli2016}, and
HH 30 (Class II) \citep{Louvet2018}, strongly support this possibility.  For
example, in TMC1A, a low-velocity and poorly collimated CO outflow was
proposed to be driven by an extended disk wind coming from the disk surface
with a radius up to 25 au from the central source.  A poorly collimated and
low-velocity rotating molecular outflow was also detected in the high-mass
source Orion BN/KL Source I \citep{Greenhill2013,Hirota2017} and it can also
be driven by an extended disk wind coming from the disk surface with a
radius up to $\sim$ 20 au from the central source.  Nonetheless, further
works are still needed to check if the rotating outflows can also be rotating
envelope material swept up by inner winds, either the wide-angle components
of X-winds or the inner disk winds.

\begin{figure}[htb]
\centering
\includegraphics[width=0.6\textwidth]
       {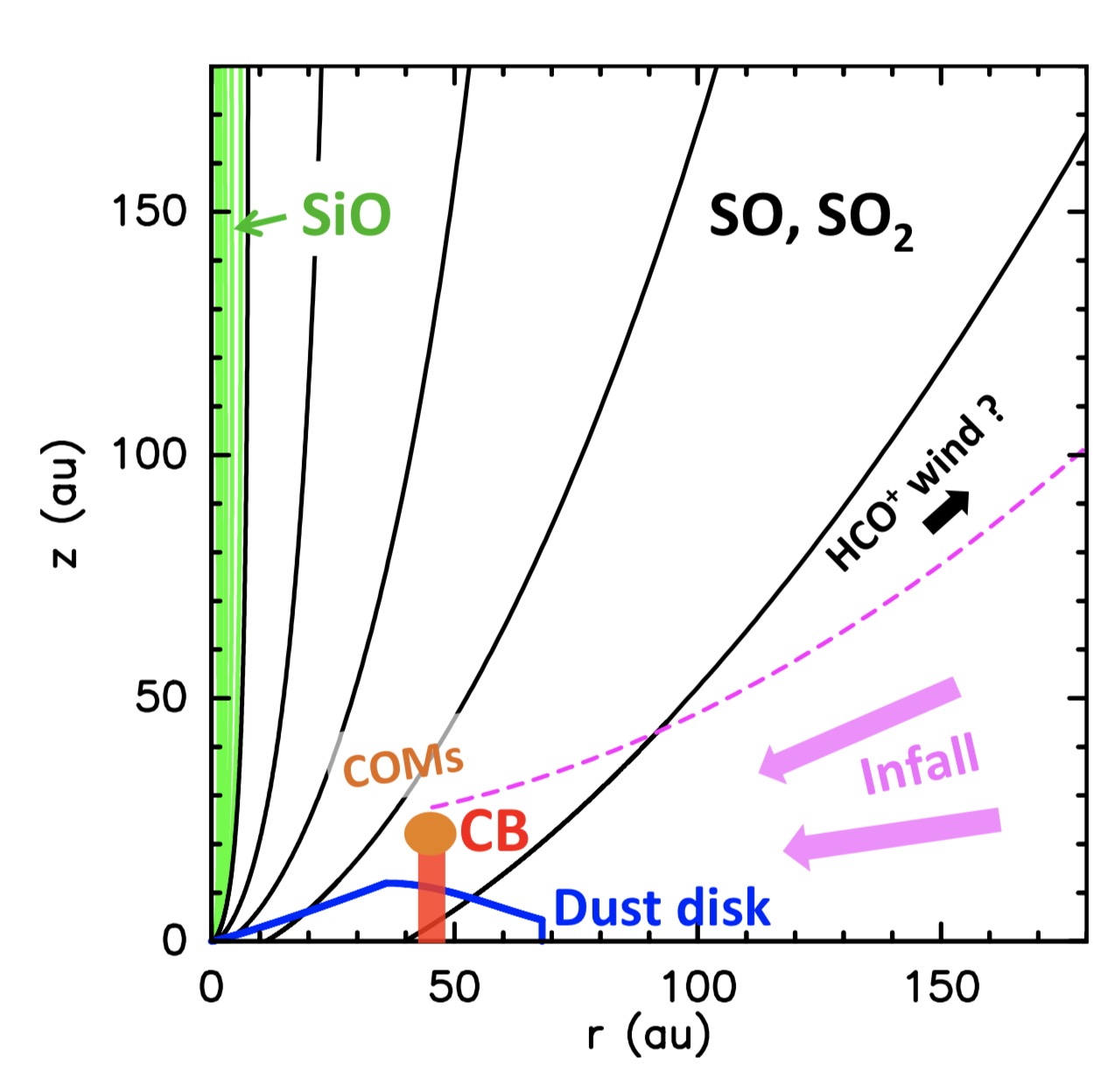}
\caption{Schematic view of the inner 180 au region of the HH212 system, showing
the SiO jet and a possible SO disk wind around the jet \citep{Tabone2017}.
Here CB means centrifugal barrier and COMs means complex organic molecules
in the disk atmosphere.
\label{fig:TaboneDwind}}
\end{figure}

\begin{figure}[htb]
\centering
\includegraphics[angle=270, width=0.8\textwidth]
       {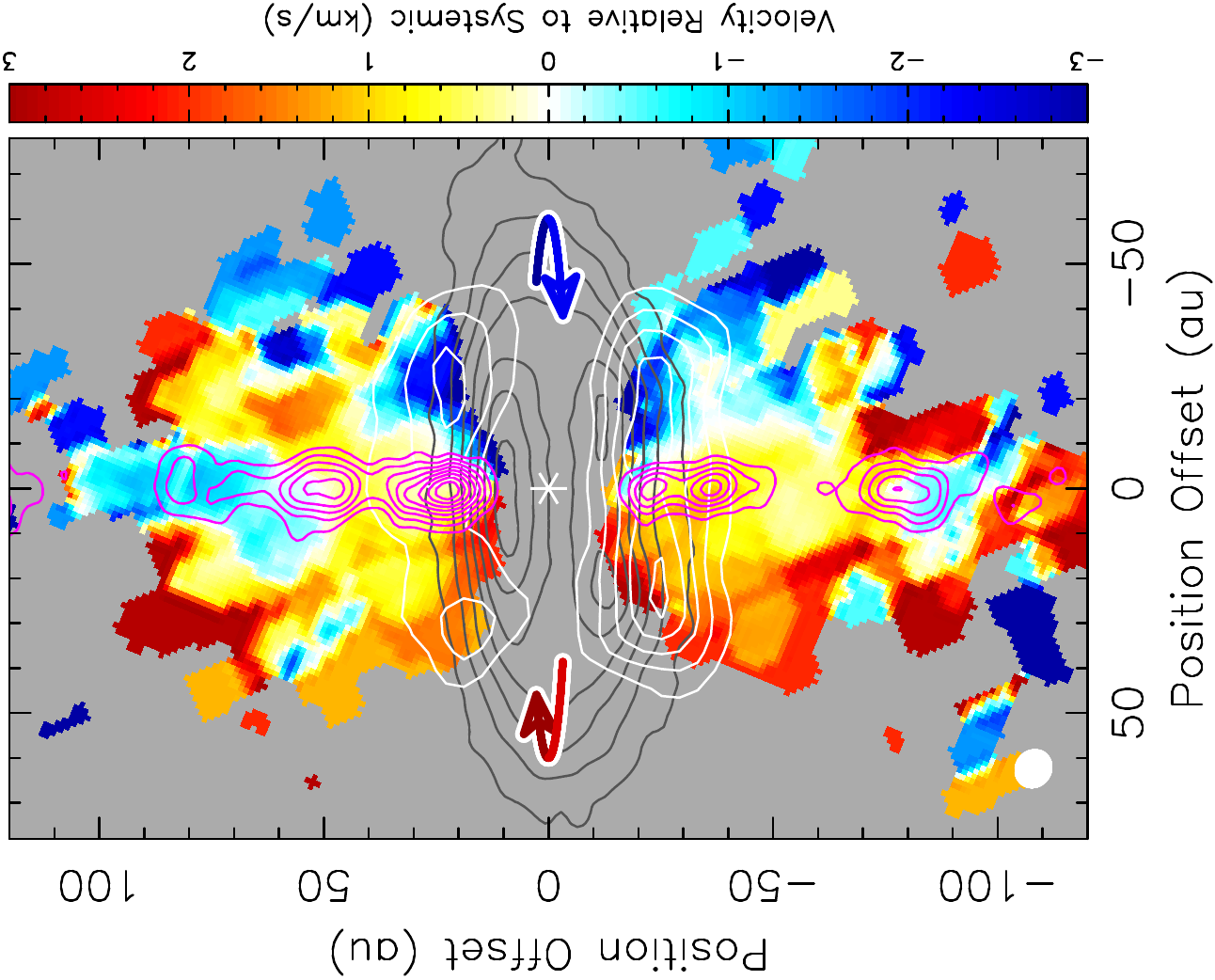}
\caption{SO rotating outflow from the disk around the SiO jet
in HH 212 \citep{Lee2018Dwind}.
Magenta contours show the SiO jet.
Gray contours show the continuum map of the disk at 850 \micron{}.
White contours show the disk atmosphere detected in CH$_2$DOH. 
Color maps show the intensity-weighted velocity of the SO outflow.
The red and blue arrows show the rotation of the disk.
\label{fig:HH212SO}}
\end{figure}

In the early phase, disk winds could also be present around the molecular
jets, e.g, in HH 212 \citep{Tabone2017,Lee2018Dwind} and HH 211
\citep{Lee2018HH211}.  For example, in HH 212, by modeling a rotating SO
outflow near the disk around the jet, \citet{Tabone2017} suggested that the
outflow can trace a wind coming from the disk at a radius up to 40 au to the
outer edge of the disk, as shown in the schematic diagram in Figure
\ref{fig:TaboneDwind}.  The disk wind could be massive and can carry away
$\sim$ 50\% of the incoming accretion flow \citep{Tabone2017}.  This SO
outflow was later observed and resolved at higher angular resolution, and
found to extend out to only $\sim$ 70 au above and below the disk midplane
\cite[][see also Fig.~\ref{fig:HH212SO}]{Lee2018Dwind}.  It has a specific
angular momentum of $\lesssim$ 40 au \vkm{}, indicating that it could trace
a disk wind launched at a radius up to $\sim$ 7 au \citep{Lee2018Dwind}. 
However, further observations are needed to check if the rotating SO outflow
could also trace the material in the rotating disk atmosphere being pushed
up and out by an inner wind launched at the innermost disk, e.g., a
wide-angle radial wind component surrounding the fast jet as in the X-wind
model.

\section{Summaries and conclusions}

Recent results of a small sample of molecular jets at high spatial and
velocity resolution have provided strong constraints on jet launching and
collimation.  The mean ratio of the mass-loss rate in the jet to the
accretion rate is $\sim$ 0.2, as expected in magneto-centrifugal
jet-launching models.  This ratio implies a magnetic lever arm parameter of
$\sim$ 5 and a jet launching radius of $\sim$ 0.04 au.  More importantly, a
clear rotation is detected in the HH 212 jet within 100 au of the central source,
with a rotation sense the same as that of the disk, confirming the role of
the jet in removing angular momentum from the disk.  The specific angular
momentum of the jet is $\sim$ 10 au \vkm{}, consistent with the jet being
launched from the innermost edge of the disk at $\sim$ 0.04 au.  The jet is
expanding roughly with the square of the distance from the central source,
roughly consistent with that predicted in magneto-centrifugal jet-launching
models, where the jet is collimated internally by its own toroidal magnetic
field.  There is also a hint of a hollow cone in this jet, because the knots
in this jet show a linear velocity structure that can come from a ring. 
Assuming that the knots are indeed rings, then the knots have a radius
slightly larger than that of a hollow cone with a launching radius of $\sim$
0.04 au, consistent with the jet tracing the innermost core of the wind in
the magneto-centrifugal jet-launching models.

 
The knots and bow shocks in the jets likely trace the internal shocks
produced by quasi-periodical variations in ejection velocity, which in turn
is induced by quasi-periodical perturbations of the accretion in the disks. 
Sideways ejections are detected in the knots and bow shocks.  Nested
internal shells are also detected, with each extending from a bow shock back
to the central source.  Up to 3 periodical variations are detected in the
jets, with one having a period of a few yrs, one with a period of a few ten
yrs, and one with a period of a few hundred yrs.  The one with a period of a
few hundred yr could be induced by gravitational instability powered by
envelope accretion.  Others could be due to binary companions, either
stellar or planetary, magneto-rotational instabilities, gravitational
instability penetrating to the inner disks, etc.

Two types of wiggles, point-symmetric and reflection-symmetric, are detected
in the jet trajectories.  The point-symmetric wiggle could be due to jet
precession, which in turn could be due to disk precession.  Disk precession can
be induced by tidal interactions in noncoplanar binary systems.  It can also
be induced by asymmetric envelope accretion, when there is a misalignment
between the flattened envelope and the disk because of a misalignment
between the magnetic axis and rotation axis in the star-forming core.  On
the other hand, the reflection-symmetric wiggle could be due to an orbital
motion of the jet sources around the binary companions.

High sensitivity observations with ALMA can detect SiO line polarization in
molecular jets due to the Goldreich--Kylafis effect, allowing us to map the
magnetic field morphology in the jets.  Recently, SiO line polarization has
been detected toward a well-defined jet within a few hundred au of the
central source, with an orientation parallel to the jet axis, indicating
that the magnetic field there could be either mainly toroidal or mainly
poloidal.  Additional polarization observations in SiO in different line
transitions are needed to resolve the ambiguity in the field morphology. 
Moreover, current-driven kink instability might have been detected in one of
the jets, suggesting a possible presence of a poloidal field in the jet core
as well.

Being launched within a radius $<$ 0.1 au of the central sources, molecular
jets are likely launched from dust-free regions.  The high content of
molecular gas in the jets likely arise because the mass-loss rate is high $\sim
10^{-6} \solarmass \mathrm{\ yr}^{-1}$, so that molecules can form quickly in the
jets.  Current observations show that a SiO jet can be detected down to within
$\sim$ 3 au of the central source, where the dynamical age is less than a
month, strongly supporting this possibility.

Rotating molecular outflows are detected around a few molecular jets,
suggestive of a presence of extended disk winds around the jets.  The
extended disk winds are expected to carry angular momentum away from the
disks, allowing the disk material to be transported to the inner parts from
the outer parts.  However, more works are needed to check if the rotating
outflows can also be rotating envelope material or disk atmosphere swept up
by inner winds, either the wide-angle components of the X-winds or the inner
disk winds.

In summary, recent observations of a small sample of molecular jets have
opened up opportunities for us to detect the jet rotation, to resolve the
jet structure and search for a hollow cone, to map the magnetic field, to
study the periodical variation in ejection and thus the periodical
perturbations in disk accretion, etc.  Future systematic observations with a
large sample of molecular jets at high spatial and velocity resolution with
ALMA are expected to lead to a breakthrough understanding in the study of jets. 
In addition, sensitive observations of the Zeeman effect in molecular lines will
allow us to derive the strength of the magnetic fields in the jets.

\begin{acknowledgements}
C.-F.L.  acknowledges grants from the Ministry of Science and Technology of
Taiwan (MoST 107-2119-M-001-040-MY3) and the Academia Sinica (Investigator
Award).  I thank the referee Bo Reipurth for his carefully reading my manuscript
and for his useful comments and suggestions. I also thank Anthony Moraghan
for his helpful suggestions on English grammar.
\end{acknowledgements}

\bibliographystyle{spbasic}      

\end{document}